%% file: main.tex
\documentclass[preprint,12pt]{elsarticle}

\usepackage[algoruled,vlined]{algorithm2e}
\usepackage{nomencl}
\usepackage{graphicx}
\usepackage{amssymb}
\usepackage{amsthm}
\usepackage{amsmath}
\usepackage{xcolor}
\usepackage{hyperref}
\usepackage{natbib}
\usepackage{xfrac}
\usepackage{arydshln}

\include{format}

\journal{arXiv}

\begin{document}

\begin{frontmatter}

\title{On modal analysis of laminated glass: Usability of simplified methods and enhanced effective thickness}

\author{Alena Zemanov\'{a}, Jan Zeman, Tom\'{a}\v{s} Janda, Jaroslav Schmidt, Michal \v{S}ejnoha}

\address{Department of Mechanics, Faculty of Civil Engineering, Czech Technical University in Prague}

\begin{abstract}
This paper focuses on the modal analysis of laminated glass beams. In these multilayer elements, the stiff glass plates are connected by compliant interlayers with frequency/temperature-dependent behavior. 
The aim of our study is
(i) to assess whether approximate techniques can accurately predict the behavior of laminated glass structures
and
(ii) to propose an easy tool for modal analysis based on the enhanced effective thickness concept by Galuppi and Royer-Carfagni.

To this purpose, we consider four approaches to the solution of the related nonlinear eigenvalue problem: a complex-eigenvalue solver based on the Newton method, the modal strain energy method, and two effective thickness concepts. A comparative study of free vibrating laminated glass beams is performed considering different geometries of cross-sections, boundary conditions, and material parameters for interlayers under two ambient temperatures. The viscoelastic response of polymer foils is represented by the generalized Maxwell model.

We show that the simplified approaches predict natural frequencies with an acceptable accuracy for most of the examples. However, there is a considerable scatter in predicted loss factors. The enhanced effective thickness approach adjusted for modal analysis leads to lower errors in both quantities compared to the other two simplified procedures, reducing the extreme error in loss factors to one half compared to the modal strain energy method or to one quarter compared to the original dynamic effective thickness method.
\end{abstract}


\end{frontmatter}

\section{Introduction}
\label{S:Intro}

Laminated glass is a multilayer composite made of glass layers and plastic interlayers, typically polymers.
These foils improve the post-fracture behavior of the originally brittle glass elements \cite{larcher2012experimental,mohagheghian2017deformation,bedon2017low}, increase their damping~\cite{Huang:2016:DMEVE,Rao:2003:RAVD}, and therefore they allow for applications prohibited to traditional glass, such as load-bearing and fail-safe transparent structures. 
Avoiding resonance and reducing noise and vibrations of laminated glass components is thus substantial not only in building structures but also in the car or ship design process and others. Thus, the reliable prediction of natural frequencies and damping characteristics associated with each vibration mode is an essential issue for the design of dynamically loaded structures.
In the case of laminated glass, free vibration analysis leads, because of the viscoelastic behavior of the foil~\citep{Andreozzi:2014:DTT}, to an eigenvalue problem with complex eigenvalues and eigenvectors, corresponding to natural angular frequencies and mode shapes. In addition, the nonlinearity due to the frequency/temperature-sensitive response of the polymer interlayer adds to the complexity of the analysis.

Several approaches to analyzing vibrations of viscoelastically damped layered composites can be found in the literature, e.g., \cite{treviso2015damping,Daya:2001:NMNEP,Huang:2016:DMEVE}.
In this paper, we broadly divide these methods into three groups: (i) numerical approaches solving the complex eigenvalue problem directly~\citep{Hamdaoui:2016:CNS}, (ii) simplified numerical approximations dealing with a real eigenmode problem only~\citep{Bilasse:2010:LNV}, and (iii)~analytical methods and effective thickness methods derived from analytical models~\cite{Aenlle:2013:FRLGE}.

The comparison of selected \textit{non-linear solvers for complex-valued problems} in~\citep{Hamdaoui:2016:CNS} shows that while most of them converge towards the same eigenvalues, their computational time and the number of iterations differ. 
The computational cost can be reduced using \textit{simplified numerical methods}, 
which deal only with a real eigenvalue problem corresponding to the delayed elasticity or take into account only the real part of the complex stiffness of the core~\citep{Bilasse:2010:LNV}. Then, the damping parameters are obtained by post-processing the real-valued eigenvalues and eigenmodes using, e.g., the modal strain energy method~\cite{Johnson:1982:FEP} to be discussed later in the paper. 

For three-layer structures with simple boundary conditions and geometry,  \textit{analytical solutions} can be derived~\citep{Mead:1969:LFRFS,Mead:1970:LFRFS,Ross:1959:DPFV}. Because of the frequency-dependent behavior of the polymer foil, they provide natural frequencies and loss factors using an iterative algorithm. 
Recently, \textit{the dynamic effective thickness approach} for laminated glass beams was proposed by~\citet{Aenlle:2013:FRLGE}, using the complex flexural stiffness introduced in~\citep{Ross:1959:DPFV}. This concept can be extended towards plates~\citep{Aenlle:2014:DETLG} and multilayer laminated glass beams~\citep{Pelayo:2017:NFADR}. The validation of this dynamic effective thickness method against the results of experimental testing in~\citep{Aenlle:2013:FRLGE,Aenlle:2014:DETLG,Pelayo:2017:NFADR} shows that  using this approach natural frequencies can be predicted with good accuracy but there is a high scatter in loss factors. 

Therefore, we want to analyze the accuracy of the response of effective thickness and other simplified approaches, to investigate their usability for the modal analysis of laminated glass elements, and to propose some improvements. 
More specifically, we
\begin{itemize}
\item perform a comparative study for free vibrating beams using selected solvers representing the three groups as introduced above and
\item propose an easy tool for modal analysis based on the enhanced effective thickness concept by~\citet{Galuppi:2012:ETL}.
\end{itemize}
To our best knowledge, no such comparison of complex and approximate models has been performed for laminated glass so far. 

All methods are compared for simply-supported, clamped-clamped, or free-free beams with symmetric and asymmetric cross-sections under different ambient temperatures. The viscoelastic behavior of polymer foil is described with the generalized Maxwell model. Several sets of parameters of the chain were taken from literature~\cite{Andreozzi:2014:DTT,Mohagheghian:2017:QSB,Shitanoki:2014:PNM} and used in our case study, to evaluate and discuss the effect of various materials used in laminated glass structures and also of different Maxwell chain parameters for the the same type of interlayer. 

The structure of this paper is as follows. The geometry of a three-layer laminated glass beam and material characterization of the glass and polymer layers is outlined in~\Sref{S:Config}. The approaches based on the finite element methods, i.e., the Newton method and the modal strain energy method, are introduced in~\Sref{S:FEMethods}. 
The closed-form formula for the complex-valued natural frequencies presented in~\Sref{S:ET} is combined with the effective thickness concept, using the dynamic effective thickness from~\cite{Aenlle:2014:DETLG} and the enhanced effective thickness~\citep{Galuppi:2012:ETL} which we adjust for modal analysis. 
The results of our case study are presented and analyzed in~\Sref{S:CS}. Finally, we summarize our findings in~\Sref{S:Conc}.

\section{Characterization of laminated glass structures}
\label{S:Config}

\subsection{Configuration of laminated glass beam}
\begin{figure}[ht]
\centerline{
	\def\svgwidth{130mm}
	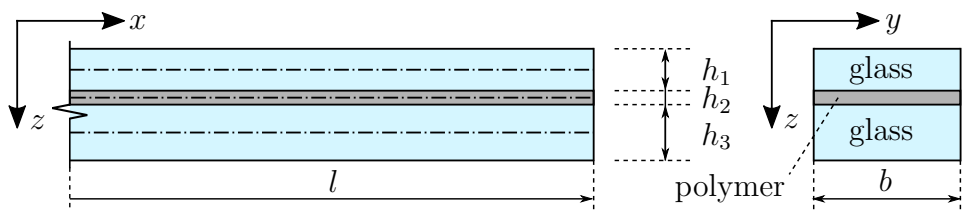
}
\caption{Basic three-layer composition of laminated glass sandwich.}
\label{fig:sandwich}
\end{figure}

In this paper, the most common three-layer configuration (with two face glass plies and one polymer interlayer, see~\Fref{fig:sandwich}) is handled for simplicity. However, the extension towards multilayer elements is possible for all discussed approaches. No slipping on the interface of a glass ply and the polymer foil is assumed. 

\subsection{Materials}
The constitutive behavior of glass and polymer layers remains the same for all presented methods. Glass is treated as an elastic material whereas the polymer behavior is assumed to be linear viscoelastic. It can be supposed that the damping of glass is small in comparison with that of the interlayer and thus it is negligible. Therefore, the behavior of a glass layer is described by three parameters: the Young modulus $\E_1$, the Poisson ratio $\PR_1$, and the density $\den_1$ (resp. $\E_3$, $\PR_3$, and $\den_3$ for the face layers of different types of glass).

The variety of interlayer materials for laminated glass is broad, ranging across the most common polyvinyl butyral (PVB), ethylene-vinyl acetate (EVA), thermoplastic polyurethane (TPU), or  the stiffest ionoplast (SGP -- SentryGlas(R) Plus). The density of the interlayer material is $\den_2$ and we assume that its Poisson's ratio $\PR_2$ is constant, see~\citep{Zemanova:2017:CVEM}.
The viscoelastic behavior of polymers is commonly described by the generalized Maxwell model or by a fractional derivative model \citep{Hamdaoui:2016:CNS}. In our study, the generalized Maxwell model is used for two reasons: (i) for laminated glass, this description is the most common in literature and (ii) according to the comparison by~\citet{Hamdaoui:2016:CNS}, it requires less computational effort than the fractional derivative model for non-linear eigenvalue problems.

\begin{figure}[ht]
\centerline{
	\def\svgwidth{100mm}
	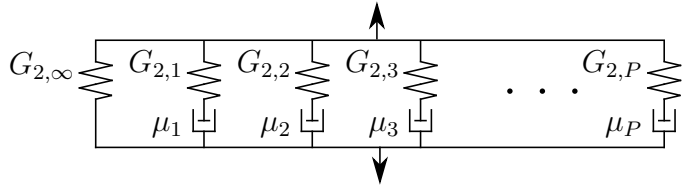
}
\caption{Generalized Maxwell chain consisting of $P$ viscoelastic units and one elastic spring.}
\label{fig:Maxwell}
\end{figure}

The schematic representation of the Maxwell model in~\Fref{fig:Maxwell} corresponds to the relaxation function provided by the Prony series, e.g.,~\cite[page 32]{Christensen:1982:TVI},
\begin{eqnarray}
\Gc(t) = \Ginf +  \sum_{\prony=1}^P \Gp \exp^{-\frac{t}{\tp}}
= \GO -  \sum_{\prony=1}^P \Gp (1-\exp^{-\frac{t}{\tp}}),
\label{eq:PronyG}					
\end{eqnarray}
where $t$ stands for the current time, $\Ginf$ is the long-term shear modulus, 
$\Gp$~denotes the shear modulus of the $p$-th unit, $\tp = \mu_{\prony} / \Gp$ its relaxation time related to the viscosity~$\mu_{\prony}$, and $\GO=\Ginf +  \sum_{\prony=1}^P \Gp$ is the elastic shear modulus of the whole chain. 

In the frequency domain, the shear modulus is given by a complex valued quantity  
\begin{equation}
\label{G_complex}
\Gc (\freqa) = 
\GO + \Gfr(\freqa),
\end{equation}
composed of the real frequency-independent part $\GO$ and the complex part dependent on the angular frequency $\freqa$ according to
\begin{equation}
\label{G_complex2}
\Gfr(\freqa)=
-\sum_\prony \Gp\frac{1}{\freqa^2 \tp^2+1} + \mathrm{i} \sum_p \Gp \frac{\freqa \tp}{\freqa^2 \tp^2+1}
.
\end{equation}
The real part (storage modulus) refers to the elastic behavior,
whereas the imaginary part (loss modulus) represents the energy dissipation effects~\citep{Christensen:1982:TVI}.
This decomposition will be useful for the formulation of the free vibration problem. Unlike in~\citep{Daya:2001:NMNEP}, the elastic shear modulus of the whole chain $\GO$ was set as the real frequency-independent part instead of the long-term shear modulus $\Ginf$ to avoid computational difficulties for chains with $\Ginf = 0$. 

\begin{figure}[ht]
\centerline{
	\includegraphics[width=110mm]{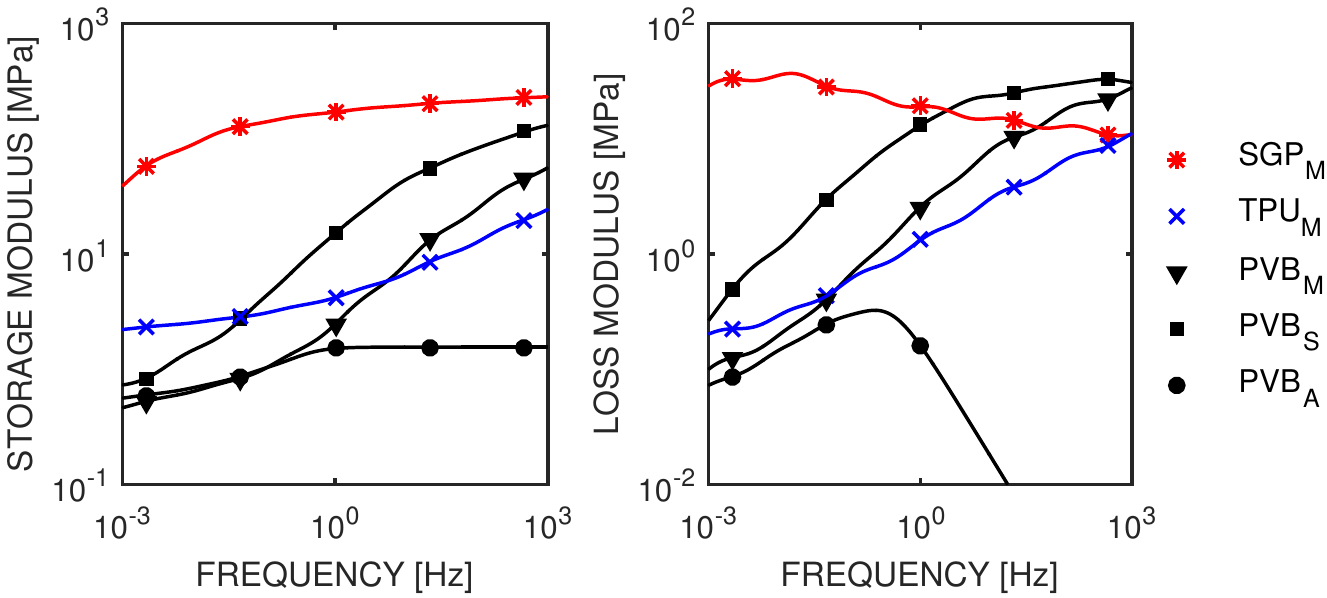}
}
\caption{The dependence of the real part (storage modulus) and the imaginary part (loss modulus) of the shear modulus $\Gc(\freqa)$ of PVB, TPU, and SGP on a real-valued frequency at $25\,^\circ$C; the Maxwell chain parameters are taken from \citep{Mohagheghian:2017:QSB} (M), \citep{Shitanoki:2014:PNM} (S), and \citep{Andreozzi:2014:DTT} (A).}
\label{fig:Modulus}
\end{figure}

The Maxwell chain parameters for different interlayer materials can be found, for instance, in \citep{Duser:1999:AGBL,Pelayo:2013:MSLGP,Biolzi:2014:LTR,Hooper:2012:MBPVB,Andreozzi:2014:DTT,Shitanoki:2014:PNM,Mohagheghian:2017:QSB}. We selected five representative sets of Prony series and related parameters, which are reported in~\ref{S:A1}. It is evident from~\Fref{fig:Modulus} that the data differ even for the same polymer type. Different test methods may result in slightly different parameters for interlayer properties, \citep{Giovanna:2016:TMD} or \citep{Andreozzi:2014:DTT}. The different content of additives and plasticizers used in the manufacturing also partially contributes to this discrepancy~\citep{Duser:1999:AGBL}. Besides, the authors mostly do not specify the frequency range for which their Prony series were determined (except for PVB$_\textrm{A}$ \cite{Andreozzi:2014:DTT}).

The effect of temperature is accounted for by using the time-temperature
superposition principle. For a given ambient temperature $T$, the relaxation times $\tp$ in~\Eref{G_complex2} are shifted by the factor $a_{T}$
derived from the Williams-Landel-Ferry equation~\cite{Williams:1955:TDR} 
\begin{equation}
\log{a_{T}} = -\frac{C_1 (T - T_0)}{C_2 + T - T_0},
\label{eq:shiftfactor}
\end{equation}
where the material constants $C_1$ and $C_2$ correspond to the reference temperature $T_0$.

\section{Finite element models}
\label{S:FEMethods}

\subsection{Refined beam element for three-layer laminated glass}

Because of the small thickness of the interlayer, we assume that the shear deformation in the viscoelastic foil is responsible for all the damping and the transverse compressive strain is negligible. Thus, we treat each layer as a one-dimensional beam element in our numerical analysis. We assume planar cross-sections of individual layers but not of the whole composite. The three layers of a laminated glass beam are constrained together with compatibility equations. This constraint can be taken into account using (i) the Lagrange multipliers -- additional unknown nodal forces holding the adjacent layers together~\cite{Zemanova:2017:CVEM}, or by (ii) the static condensation of the dependent generalized displacements~\cite{rikards1993finite}. The second option is used in this study, because no delamination is assumed during the modal analysis. 

\begin{figure}[ht]
\centerline{
	\def\svgwidth{130mm}
	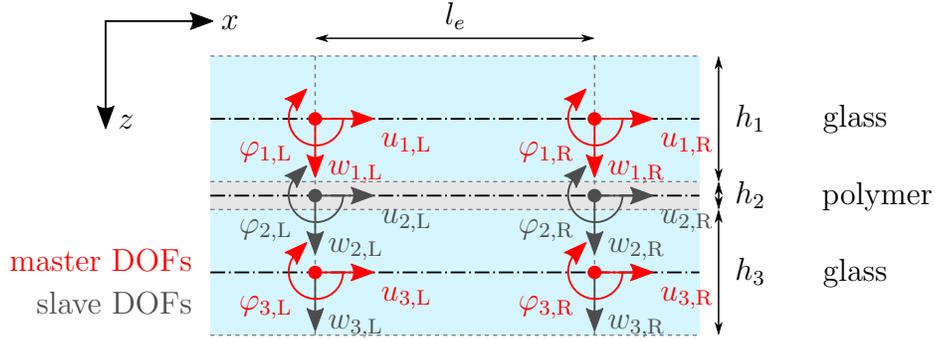
}
\caption{Master and slave degrees of freedom (DOFs) of the refined beam element: horizontal central displacements $\uL{\bullet}$, $\uP{\bullet}$, deflections $\wL{\bullet}$, $\wP{\bullet}$, and rotations $\rotL{\bullet}$, $\rotP{\bullet}$.}
\label{fig:superelement}
\end{figure}

According to the Timoshenko beam theory, there are nine unknowns per cross-section of a three-layer beam -- the horizontal and vertical centerline displacements and the rotations for each layer, see~\Fref{fig:superelement}. Four unknowns can be eliminated using four inter-layer compatibility conditions assuming the perfect adhesion in the horizontal and vertical directions. This elimination together with the outline of the element stiffness and mass matrices can be found in~\ref{S:A_NA}. Note that we use linear basis functions for the evaluation of stiffness and mass matrices, and selective integration scheme to avoid shear locking. 

\subsection{Governing equations for nonlinear eigenvalue problem}

After the discretization, the free vibration problem is described by the governing equation, e.g.,~\citep{Daya:2001:NMNEP},
\begin{equation}
\label{eq:CGEQ1}
\left( \MK (\freqa) - \freqa^2 \MM \right) \Memode = \M{0}.
\end{equation}
The eigenvalues and the eigenvectors solving the problem are complex-valued and represent the squared angular frequencies $\freqa$ and the associated mode shapes $\Memode$. 
The mass matrix $\MM$ is real-valued and constant, frequency independent, whereas the complex stiffness matrix $\MK$ depends on the complex frequency $\freqa$ of the system. Similarly to \Eref{G_complex}, the stiffness matrix is decomposed as
\begin{equation}
\label{eq:stiffness}
\MK (\freqa) = \MKO + \MKfr(\freqa).
\end{equation}
The elastic matrix $\MKO$ includes the contributions of the glass faces and the stiffness matrix of the interlayer corresponding to the instantaneous shear modulus $G_0$. 
The frequency dependent part $\MKfr(\freqa) = \Gfr(\freqa) \MKconst$, where $\MKconst$ is a constant matrix.

Then, the free vibration problem can be rewritten as 
\begin{subequations}
\label{eq:SCGEQ}
\begin{align}
\left( \MKO +  \Gfr (\freqa)\MKconst - \freqa^2 \MM \right) \Memode = \M{0},
\label{eq:CGEQ2}
\\
\Memode_{\rO}\trn (\Memode - \Memode_{\rO}) = {0}, 
\label{eq:AEQ}
\end{align}
\end{subequations}
where \Eref{eq:AEQ} is introduced to obtain a well-posed problem, see \citep{Daya:2001:NMNEP}.
The mode shape $\Memode_{0}$ is the eigenvector solving the real-eigenvalue problem
\begin{equation}
\label{eq:RGEQ}
\left( \MKO - \freqa_{\rO}^2 \MM \right) \Memode_{\rO} = \M{0}.
\end{equation}
No damping is thus assumed in this auxiliary problem, and the stiffness matrix is constant (frequency-independent). To solve~\Eref{eq:RGEQ}, we used a solver based on the implicitly restarted Arnoldi method~\citep{radke1996matlab}.

\subsection{Natural frequencies and loss factors}
The natural frequency $\freq$ and the modal loss factor $\loss$ associated with the same mode shape are determined from the relevant squared frequencies $\freqa^2$ from~\Eref{eq:CGEQ2} according to, e.g.,~\citep{Daya:2001:NMNEP},
\begin{equation}
\label{eq:fHzLoss1}
\freqa^2 = (2\pi\freq)^2 \left(1 +  \mathrm{i} \loss \right),
\end{equation}
thus
\begin{subequations} 
\label{eq:fHzLoss2}
\begin{align}
\freq &= \frac{\sqrt{\textrm{Re}[\freqa^2]}}{2\pi}, \\
\loss &= \frac{\textrm{Im}[\freqa^2]}{\textrm{Re}[\freqa^2]}.
\end{align}
\end{subequations}
Note that we decided to omit the index referring to the given mode shape to avoid a profusion of notation. Later in~\Sref{S:CS}, we compare the natural frequencies and loss factors corresponding to the first three mode shapes.

\subsection{Complex-eigenvalue solver using Newton method (CNM)}
\label{S:Newton}
The Newton method is applied to the extended system~\eqref{eq:SCGEQ} to obtain the complex-valued pair of an eigenvalue and an eigenvector. Our approach belongs to the class of methods iterating each individual eigenpair independently, starting from the initial eigenpairs solving~\Eref{eq:RGEQ}. We express the $(k+1)$-th approximations of the searched frequency and mode shape from the previous $k$-th approximations and their increments in the form
\begin{subequations}
\begin{align}
\ite{\step+1}\freqa &= \ite{\step}\freqa + \delta \freqa,
\\
\ite{\step+1}\Memode &= \ite{\step}\Memode + \delta \Memode. 
\end{align}
\end{subequations}
After linearization and evaluation of the Jacobian matrix, the Newton method leads to the system of linear equations, e.g., \citep[page~73, Eq.~(4.8)]{Schreiber:2008:PhD},
\begin{align}\label{eq:system}
\begin{bmatrix}
\MT (\ite{\step}\freqa)
& 
\ppd{\MT}{\freqa} (\ite{\step}\freqa)
\ite{\step}\Memode
\\
\Memode_{\rO}\trn 
& 
\M{0}
\end{bmatrix}
\begin{bmatrix}
\ite{\step+1}\Memode
\\
\delta \freqa
\end{bmatrix}
=
\begin{bmatrix}
\M{0}
\\
\Memode_{\rO}\trn \Memode_{\rO}
\end{bmatrix},
\end{align}
where the operator $\MT$ and its derivative are provided by 
\begin{subequations}
\begin{align}
\MT (\freqa) & = \MKO + 
\Gfr (\freqa)\MKconst - \freqa^2 \MM,
\\
\ppd{\MT(\freqa)}{\freqa}
& =
\ppd{\Gfr(\freqa)}{\freqa}\MKconst - 2 \freqa \MM.
\end{align}
\end{subequations}

The initial pair $\ite{\step = 0}\freqa = \freqa_{\rO}$ and $\ite{\step = 0}\Memode = \Memode_{\rO}$ is obtained from the real-eigenvalue problem in~\Eref{eq:RGEQ} and the stopping criterion is defined by
\begin{equation}
\frac{\|\MT (\ite{\step}\freqa)\ite{\step}\Memode\|}{\| \ite{\step}\Memode\|} < \epsilon_\mathrm{tol}
\end{equation}
with the norm of residual weighted by the norm of the relevant mode shape and the user defined tolerance $\epsilon_\mathrm{tol}$.

Our proposed complex-valued eigenvalue solver was verified against the semi-analytical model by~\citet{mead2007measurement}; the results for this model were adopted from~\citep{Aenlle:2013:FRLGE} for a laminated glass beam with free ends. The response is in a good agreement; the errors in frequencies are below 0.5\%, and in loss factors are less than 2\%. These minor discrepancies can be caused by different assumptions for the semi-analytical model (e.g., zero Young's modulus for the interlayer), by a different stopping tolerance of iterative solvers (which is not specified in~\citep{Aenlle:2013:FRLGE}), or by the ambiguity in Prony series, where the conversion of the Young modulus to the shear modulus is unclear (the assumption of a constant value of the Poisson ratio vs. the bulk modulus, see~\citep{Pelayo:2013:MSLGP,Zemanova:2017:CVEM}).

\subsection{Real-eigenvalue solver and modal strain energy method (MSE)}
\label{S:MSE}

Especially for large-scale structures, the direct evaluation of the complex-valued solution can be expensive. Therefore, the natural frequencies are often approximated from a real-valued simplification of the complex-valued problem, recall~\Eref{eq:SCGEQ}, as
\begin{subequations}
\label{eq:SRGEQ}
\begin{align}
\left( 
\MK_\rAp(\ite{\step}\freqa_\rAp)
- \ite{\step+1}\freqa_\rAp^2 \MM \right) \ite{\step+1}\Memode_\rAp = \M{0},
\label{eq:CREQ2}
\\
\Memode_{\rO}\trn (\ite{\step+1}\Memode_\rAp - \Memode_{\rO}) = {0}.
\label{eq:AREQ}
\end{align}
\end{subequations}
Due to the real-valued approximation of the stiffness matrix $\MK_\rAp(\ite{\step}\freqa_\rAp)$, the $({\step+1})$-th estimates of the eigenpairs are real squared angular frequencies $\ite{\step+1}\freqa_\rAp^2$ and real mode shape vectors $\ite{\step+1}\Memode_\rAp$.
%
We consider that the approximate stiffness matrix\footnote{Different ways for determining the real-valued approximate stiffness matrix $\MK_\rAp$ can be found in the literature, see~\citep{Johnson:1982:FEP,Bilasse:2010:LNV}.
} 
corresponds to the real part of the whole stiffness matrix evaluated for the relevant frequency
\begin{equation}
\MK_\rAp(\freqa_\rAp) = \MKO +  \textrm{Re}[\Gfr (\freqa_\rAp)]\MKconst.
\end{equation}
Therefore, the stiffness matrix is updated iteratively until the convergence of the angular frequency
\begin{equation}
\label{eq:conv_freq}
\frac{\| \ite{\step+1}\freqa_\rAp - \ite{\step}\freqa_\rAp \|}{\| \ite{\step+1}\freqa_\rAp \|} < \epsilon_\mathrm{tol}.
\end{equation}

To estimate the loss factors from the converged real eigenpairs ($\ite{\step+1}\freqa_\rAp$, $\ite{\step+1}\Memode_\rAp$), the modal strain energy method was introduced by~\citet{Ungar:1962:LFVS} and later popularized for finite element models by~\citet{Johnson:1982:FEP}.
This method assumes that no changes occur in the damped mode shapes,~\Eref{eq:MSE_loss}, and therefore it is suitable only for lightly damped structures~\cite{treviso2015damping}.
The modal loss factor of each individual mode is determined according to
\begin{equation}
\loss_\rAp = \frac{\ite{\step+1}\Memode_\rAp\trn 
\textrm{Im}[\Gfr (\ite{\step+1}\freqa_\rAp)]\MKconst
\ite{\step+1}\Memode_\rAp}{\ite{\step+1}\Memode_\rAp\trn \MK_\rAp (\ite{\step+1}\freqa_\rAp) \ite{\step+1}\Memode_\rAp}.
\label{eq:MSE_loss}
\end{equation}
The derivation of this formula from Rayleigh's quotient is described in~\citep{Johnson:1982:FEP}. 

This iterative procedure provides more accurate values of the natural frequencies and the loss factors than the original approximation technique in~\citep{Johnson:1982:FEP}. The original one starts from a constant initial value of the interlayer shear modulus and then only adjusts the computed loss factor by a correction factor taking into account the change of material properties due to the frequency shift. 

\section{Effective thickness approaches}
\label{S:ET}
\subsection{Closed-form expression for natural angular frequencies of beams}
A few effective thickness formulations can be found in the literature for laminated glass beams and plates under static loading, whereas, to the best of our knowledge, only one effective thickness approach exists for dynamic problems~\citep{Aenlle:2013:FRLGE,Aenlle:2014:DETLG}. In general, the effective thickness methods are based on calculating a constant thickness of a monolithic element (with the same width and length), which gives the same response as the laminated glass beam under the identical loading and boundary conditions. For example, a deflection-effective thickness and stress-effective thicknesses~\citep{bennison2008high,Galuppi:2012:ETL} are defined for laminated glass structures under static bending to obtain the same extreme values of deflection and stresses. 

In this paper, this effective thickness concept is applied to vibrating laminated glass beams, and the dynamic effective thickness is used for the modal parameters calculation. The analytical expressions  for the natural angular frequencies for a monolithic beam are then given by, e.g.,~\citep[Eq. (22)]{Aenlle:2013:FRLGE},
\begin{equation}
\label{eq:freqEff}
\ite{\step+1}\freqa_\rE^2
= \frac
{\wn^4 \E_1 h_\rE^3(\ite{\step}\freqa_\rE)}{12\bar{m}},
\end{equation}
where $\wn$ is the wavenumber, $\E_1$ is the Young modulus of glass, $\bar{m}$ is the mass per unit length and width of the beam ($\bar{m} = \den_1 h_1 + \den_2 h_2 + \den_3 h_3$), and $h_\rE(\ite{\step}\freqa_\rE)$ is the dynamic effective thickness. Two expressions for this effective thickness will be introduced in the next two sections. Note that the wavenumbers are usually expressed in the form $\wn=\hat{\wn}/l$, where $l$ is the length of the beam and $\hat{\wn}$ depends on boundary conditions and a mode shape~\citep[Chapter~18]{clough2003dynamics}. 

Due to the dependency of the dynamic effective thickness on frequency, the search for $\freqa_\rE^2$ requires iterations.
When the convergence is achieved according to the criterion~\eqref{eq:conv_freq}, the frequency $\freq$ and the loss factor $\loss$ are determined from the complex valued $\freqa_\rE^2$ from~Eqs.~(\ref{eq:fHzLoss1}--\ref{eq:fHzLoss2}).

\subsection{Dynamic effective thickness concept (DET)}
\label{S:DET}
The dynamic effective thickness was introduced by~\citet{Aenlle:2013:FRLGE}, and the iterative algorithm was extended to plates in~\citep{Aenlle:2014:DETLG}.
This complex-valued effective thickness was derived from the closed-form formula for effective stiffness by~\citet{Ross:1959:DPFV} for a three-layer, simply-supported beam with purely elastic face layers and a linearly viscoelastic core. However, this analytical model is also used for different boundary conditions using the known relevant wavenumbers, e.g., for free-free beams in~\citep{Aenlle:2014:DETLG}.

Under the assumption that the glass parameters are the same for both glass layers ($\E_3=\E_1$, $\PR_3=\PR_1$, and $\den_3=\den_1$) the expression for the dynamic effective thickness from~\citep{Aenlle:2014:DETLG} holds
\begin{equation}
\label{eq:h_ef}
h_\rE(\freqa_\rE)=
\sqrt[3]{\left(h_1^3+h_3^3\right)
\left(
1 + Y 
\left(
1 + \frac{h_1}{g(\freqa_\rE) (h_1+h_3)}
\right)^{-1}
\right)
},
\end{equation}
where the geometric parameter
\begin{equation}
\label{eq:Y_par}
Y = \frac{12 h_1 h_3 (0.5 h_1+h_2+0.5 h_3)}{(h_1^3+h_3^3)(h_1+h_3)}
\end{equation}
depends on the thicknesses of individual layers, and the shear parameter
\begin{equation}
\label{eq:g_par}
g (\freqa_\rE) = \frac{\Gc (\freqa_\rE)}{\E_1 h_3 h_2 \wn^2} 
\end{equation}
additionally includes the wavenumber $\wn$, the Young modulus of glass $\E_1$, and the complex shear modulus of polymer $\Gc (\freqa_\rE)$; recall \Eref{G_complex}. 
 
 \subsection{Enhanced effective thickness adjusted for modal analysis (EET)}
 \label{S:SET}
One of the effective thickness approaches for layered beams under static loading is the enhanced effective thickness method by~\citet{Galuppi:2012:ETL} or~\citet{Galuppi:2012:PEFD}. It is derived from an energy-based variational formulation. Under the same assumption as in the previous section ($\E_3=\E_1$, $\PR_3=\PR_1$, and $\den_3=\den_1$), the enhanced effective thickness for the deflection is then expressed as~\citep[Eq. (4.14)]{Galuppi:2012:ETL},
\begin{equation}
h_\rE=\sqrt[3]{\frac{1}{\displaystyle \frac{\zeta}{h_1^3+h_3^3+12I_s}+\displaystyle \frac{1-\zeta}{h_1^3+h_3^3}}},
\label{eq:EET_hef}
\end{equation}
with the coefficient of shear cohesion $\zeta$, the non-dimensional ratio of the glass and interlayer stiffnesses $\mu$, and the shape coefficient $\psi$ dependent on a normalized shape of the deflection curve $g(x)$ of a homogeneous beam. These parameters follow from
\begin{align}\label{eq:EET}
\zeta=\frac{1}{1+\displaystyle \frac{I_1+I_3}{\mu I_\mathrm{tot}}\frac{A_1A_3}{A_1+A_3}\psi}, 
&&
\mu=\frac{\Gc b}{E_1 h_2},
&&
\psi=\frac{\int_0^l{\left(g''(x)\right)^2}\mathrm{d} x}{\int_0^l{\left(g'(x)\right)^2}\mathrm{d}x}.
\end{align}
The cross-section areas $A_i$ for $i={1,2,3}$ and the second moments of area $I_i$, $I_\mathrm{tot}$, and $I_\mathrm{s}$ are defined by relations, recall~\Fref{fig:sandwich}, 
\begin{subequations}
	\begin{align}
	I_i & =\frac{1}{12}bh_i^3,
	\quad  
	A_i =bh_i,
	\\
	I_\mathrm{tot} & =I_1+I_3+\frac{A_1A_3}{A_1+A_3}(h_2+0.5(h_1+h_3))^2,
	\\
	I_\mathrm{s} & = \frac{h_1h_3}{h_1+h_3}(h_2+0.5(h_1+h_3))^2.
	\end{align}
\end{subequations}

Two intuitive adjustments of this method were made for its use in modal analysis. In~\Eref{eq:EET}, the complex-valued shear modulus for interlayer from \Eref{G_complex} was used for coefficient $\mu$, and the shape function of the deflection under static loading $g$ was replaced by the one corresponding to the $n$-th mode shape of a monolithic beam under given boundary conditions, e.g., \cite[Chapter~18]{clough2003dynamics}. The resulting parameters $\psi$, corresponding to three basic boundary conditions, are summarized in~\Tref{tab:psi}.
\begin{table}[h]
\centering
\begin{tabular}{lccc}
\hline
Beam & \multicolumn{3}{c}{Mode}\\
& 1 & 2 & 3
\\
\hline
simply-supported & ${\pi^2}/{l^2}$ & ${(2\pi)^2}/{l^2}$ & ${(3\pi)^2}/{l^2}$\\
fixed-fixed & ${40.7}/{l^2}$ & ${82.6}/{l^2}$ & ${148}/{l^2}$\\
free-free & ${10.1}/{l^2}$ & ${34.9}/{l^2}$ & ${78.2}/{l^2}$
\\
\hline
\end{tabular}
\caption{Shape coefficients $\psi$ for three basic boundary conditions and the first three mode shapes for a beam of the length $l$.}
\label{tab:psi}
\end{table}

This effective thickness is also complex-valued, as in the DET approach in the previous section, and the natural frequencies and the damping are evaluated according to~\Eref{eq:fHzLoss2}. Note that this procedure can be extended to the plate structures, similarly to its static variant~\cite{Galuppi:2012:ETLGP,Galuppi:2012:PEFD}, but we leave this to future work.

\section{Case study}
\label{S:CS}

In this section, the usability of the three methods introduced above is assessed for laminated glass beams. The section is divided into two parts: the first introducing the selected test examples and the second discussing the results and the effect of input data on the usability of the modal strain energy method and the two effective thickness approaches.

\subsection{Examples}

We examined a collection of 63 examples, which results from all combinations of input data for simply-supported (S-S), free-free (F-F), and clamped-clamped (C-C) beams in~\Tref{tab:Input}. 
 \begin{table}[h]
\centering
\begin{tabular}{l c l}
\hline
Attribute & Options/Values & Units\\
\hline
supports & S-S, F-F, C-C & \\
$l$ $\times$ $b$ & 1 $\times$ 0.1 & m $\times$ m\\
$h_1/h_2/h_3$ & 10/0.76/10, 15/0.76/5, 10/1.52/10 & mm/mm/mm\\
interlayer & SGP$_\textmd{M}$, TPU$_\textmd{M}$, PVB$_\textmd{M}$, PVB$_\textmd{S}$,
PVB$_\textmd{A}$ &\\
temperature & 25, 50 (for PVB$_\textmd{S}$ and PVB$_\textmd{A}$) & $^\circ\textrm{C}$\\
\hline
\end{tabular}
\caption{Input data and boundary conditions.}
\label{tab:Input}
\end{table}

The length $l$ and the width $b$ of the laminated beam were constant, whereas three cross-section configurations $h_1/h_2/h_3$ were considered to assess the effect of symmetry/asymmetry of the cross-section layout and the thickness of the interlayer, see~\Fref{fig:sandwich}.

Material properties of glass and three interlayer materials from three different sources, \Fref{fig:Modulus}, are summarized in~\ref{S:A1}. The calculations were carried out for two ambient temperatures: the room temperature 25$\,^\circ\textrm{C}$ for all cases and the elevated temperature 50$\,^\circ\textrm{C}$ for two PVB-interlayers, where the temperature-shift parameters were specified.

Finally, for the sake of completeness, \Tref{tab:wn} contains the relation for wavenumbers which we used in effective thickness approaches.
\begin{table}[h]
\centering
\begin{tabular}{lccc}
\hline
Beam & \multicolumn{3}{c}{Mode}\\
& 1 & 2 & 3
\\
\hline
simply-supported & ${\pi}/{l}$ & ${2\pi}/{l}$ & ${3\pi}/{l}$\\
fixed-fixed or free-free & ${4.7300}/{l}$ & ${7.8532}/{l}$ & ${10.996}/{l}$
\\
\hline
\end{tabular}
\caption{Wavenumbers $\wn$ for three basic boundary conditions and first three mode shapes for a beam of the length $l$.}
\label{tab:wn}
\end{table}

 \subsection{Results and discussion}
\label{S:Comp}

The natural frequencies and loss factors for the first three modes\footnote{For F-F beam the first three modes corresponding to the rigid body motion were skipped in all solvers and are not considered in this comparison.} were determined according to the detailed and simplified algorithms described in Sections~\ref{S:FEMethods} and \ref{S:ET}.
We set the tolerance $\epsilon_\mathrm{tol}$ to $10^{-5}$ for all methods.
For finite element solvers, each beam layer was discretized by 200 elements. Compared to the results obtained by using a refined discretization by 300 elements per length, the largest errors for all tested examples is below 0.03\% for natural frequencies and below 0.8\% for loss factors. 

The modal response obtained by the simplified methods is compared against the reference method based on the complex-valued solver using the Newton method, recall~\Sref{S:Newton}. Each method is discussed in a separate section. 

\subsubsection{MSE against CNM}
\label{S:CompMSE}

\begin{figure}[p]
\centerline{
	\includegraphics[width=110mm]{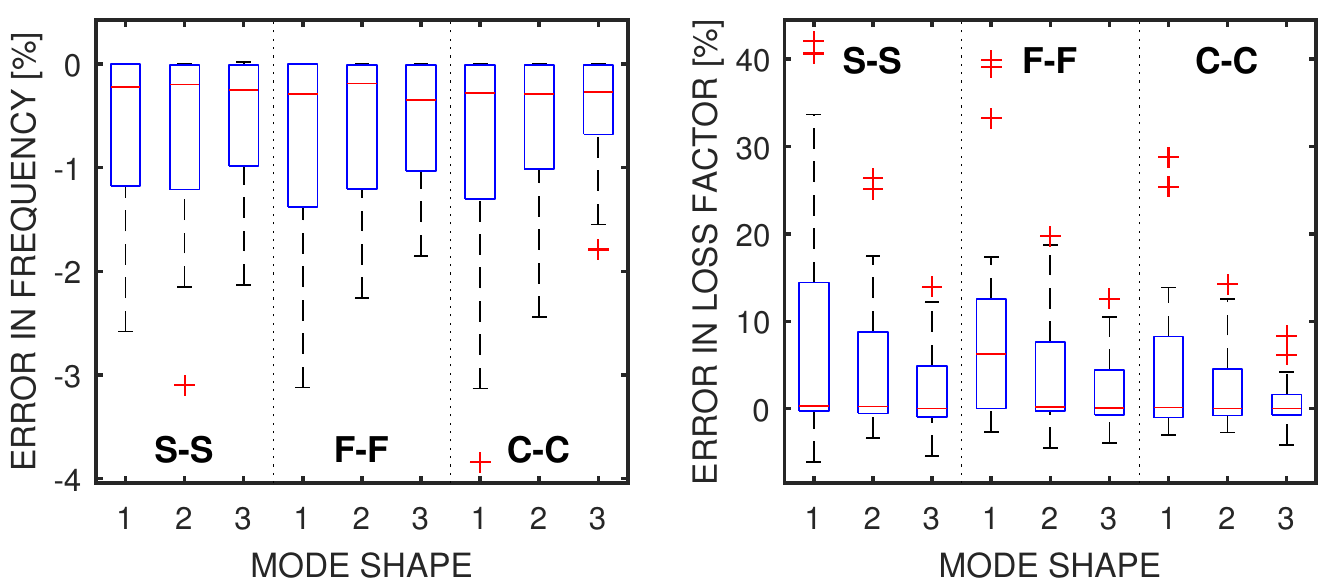}
}
\caption{Errors in natural frequencies and loss factor by the modal strain energy method (MSE) against the reference solution (CNM) for first three mode shapes for simply-supported (SS), free-free (F-F), and clamped-clamped (C-C) beams.}
\label{fig:e3MSE}
\end{figure}

The box-plots in~\Fref{fig:e3MSE} visualize the errors of natural frequencies and loss factors obtained using the MSE method against the results of the reference method (CNM). The red mark inside the box indicates the median, the bottom and top edges of the box indicate the 25th and 75th percentiles, respectively. The boxplot whiskers have the standard maximum length of 1.5 times the interquartile range, and the remaining data points are outliers plotted individually.

\begin{figure}[p]
\centerline{
\includegraphics[width=110mm]{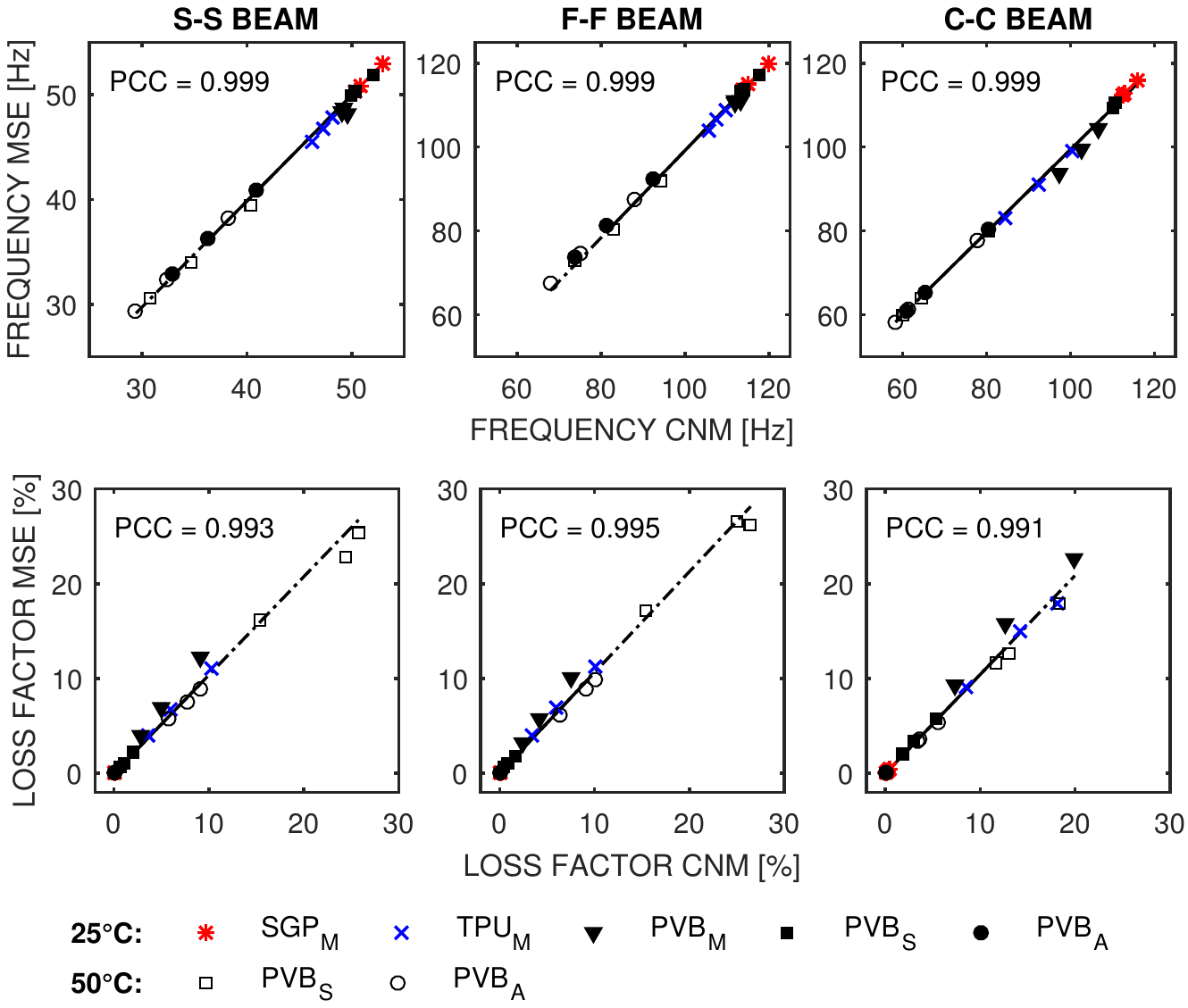}
}
\caption{Quantile-quantile plot for the first mode shape for natural frequencies and loss factors. The response by the simplified method (MSE) is plotted against the reference solution (CNM) along with the Pearson correlation coefficient (PCC).  The Maxwell chain parameters for PVB, TPU, and SGP are taken from~\citep{Mohagheghian:2017:QSB} (M), \citep{Shitanoki:2014:PNM} (S), and \citep{Andreozzi:2014:DTT} (A).}
\label{fig:QQ_MSE}
\end{figure}

For natural frequencies, the error for all 63 examples is less than 4$\%$. More specifically, the error is less than 1.5$\%$ in 75$\%$ of the tested cases and the average error is less than 0.5$\%$. Thus, this solver provide a reasonable approximation of natural frequencies, which are mostly sufficiently accurate for design purposes. There is no substantial difference in errors for the three tested boundary conditions, but the errors slightly decrease for higher mode shapes.

However, the loss factors differ significantly from the reference solution. The errors stay below 15$\%$ for 75$\%$ of configurations, but they increase up to 42$\%$ in some cases. The error decreases with an increasing number of kinematic boundary constraints or for higher modes. The highest errors in loss factors correspond to the first mode shape; the errors decrease to about 25$\%$ for the second and below 15$\%$ for the third mode shape.

For further consideration, the quantile-quantile plots in~\Fref{fig:QQ_MSE} show the values of natural frequencies and loss factors by the simplified method (MSE) against the reference solution (CNM) corresponding to the first mode shape for all cases from~\Tref{tab:Input}. These values of frequencies and loss factors are strongly influenced by the effect of temperature and the interlayer type. Also, the highest errors for loss factors up to 42$\%$ appears only for the PVB$_\textrm{M}$ foil; the errors remain below 18$\%$ for the other foils. The samples with the PVB$_\textrm{A}$ foil show entirely different response than those with the other two PVB foils because this the corresponding eigenfrequencies fall outside the frequency range of this Prony series~\cite{Andreozzi:2014:DTT}.

\subsubsection{DET against CNM}
\label{S:CompDET}

\begin{figure}[p]
\centerline{
	\includegraphics[width=110mm]{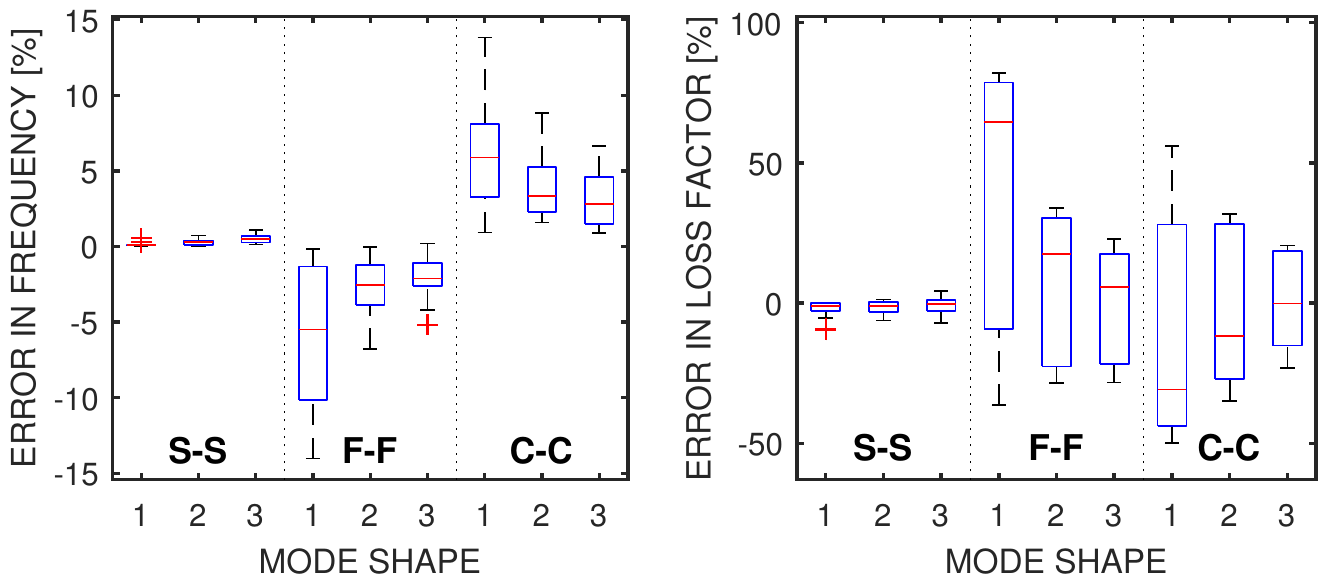}
}
\caption{Errors in natural frequencies and loss factor by the dynamic effective thickness method (DET) against the reference solution (CNM) for first three mode shapes for simply-supported (SS), free-free (F-F), and clamped-clamped (C-C) beams.}
\label{fig:e3DET}
\end{figure}

\begin{figure}[p]
\centerline{
\includegraphics[width=110mm]{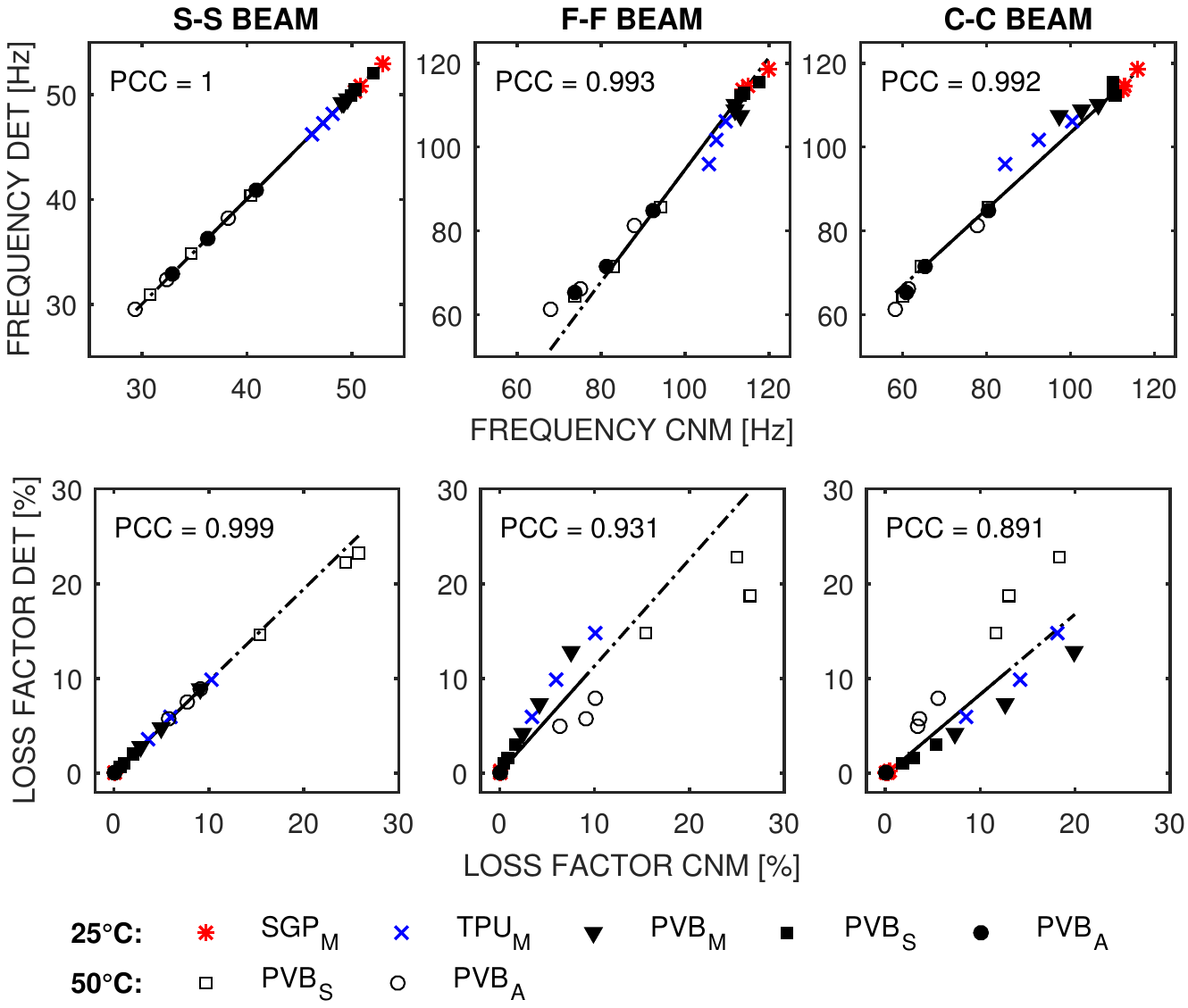}
}
\caption{Quantile-quantile plot for the first mode shape for natural frequencies and loss factors. The response by the dynamic effective thickness method (DET) is plotted against the reference solution (CNM) along with the Pearson correlation coefficient (PCC).  The Maxwell chain parameters for PVB, TPU, and SGP are taken from \citep{Mohagheghian:2017:QSB} (M), \citep{Shitanoki:2014:PNM} (S), and \citep{Andreozzi:2014:DTT} (A).}
\label{fig:QQ_DET}
\end{figure}

The comparison of the modal response by the dynamic effective thickness method (DET) against that by the reference method (CNM) are, similarly to the previous section, shown in terms of errors of natural frequencies and loss factors in~\Fref{fig:e3DET} and quantile-quantile plots in~\Fref{fig:QQ_DET}.


It is clear that this method gives very accurate results for the simply-supported beams (SS). The errors are below 1$\%$ in frequencies, and the loss factor predictions are quite accurate as well with errors below 10$\%$. For the other boundary conditions, the method does not provide satisfactory approximations, especially for loss factors, even if we used an adjusted wavenumber $k$ for given boundary conditions. This can be also seen in the quantile-quantile plot in~\Fref{fig:QQ_DET}. The errors in frequencies do not exceed 15$\%$ (the 75th percentile is below 10$\%$) and the errors in loss factors 85$\%$. The errors are decreasing for the higher mode shapes.

\subsubsection{EET against CNM}
\label{S:CompSET}

\begin{figure}[p]
\centerline{
	\includegraphics[width=110mm]{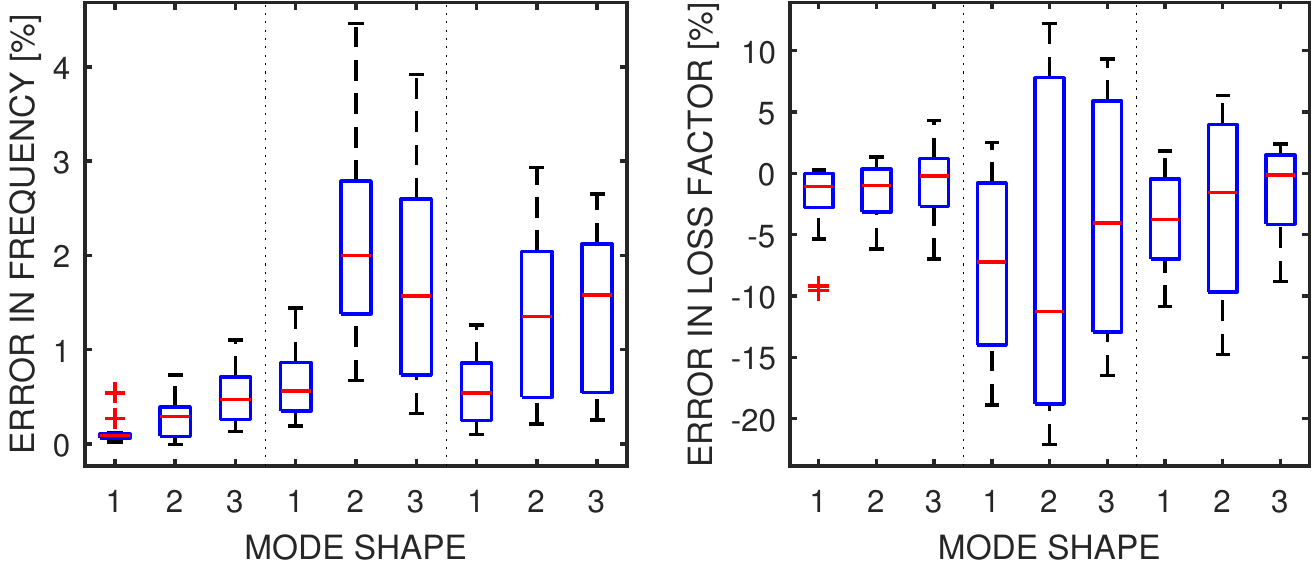}
}
\caption{Errors in natural frequencies and loss factor by the adjusted enhanced effective thickness method (EET) against the reference solution (CNM) for first three mode shapes for simply-supported (SS), free-free (F-F), and clamped-clamped (C-C) beams.}
\label{fig:e3SET}
\end{figure}

\begin{figure}[p]
\centerline{
\includegraphics[width=110mm]{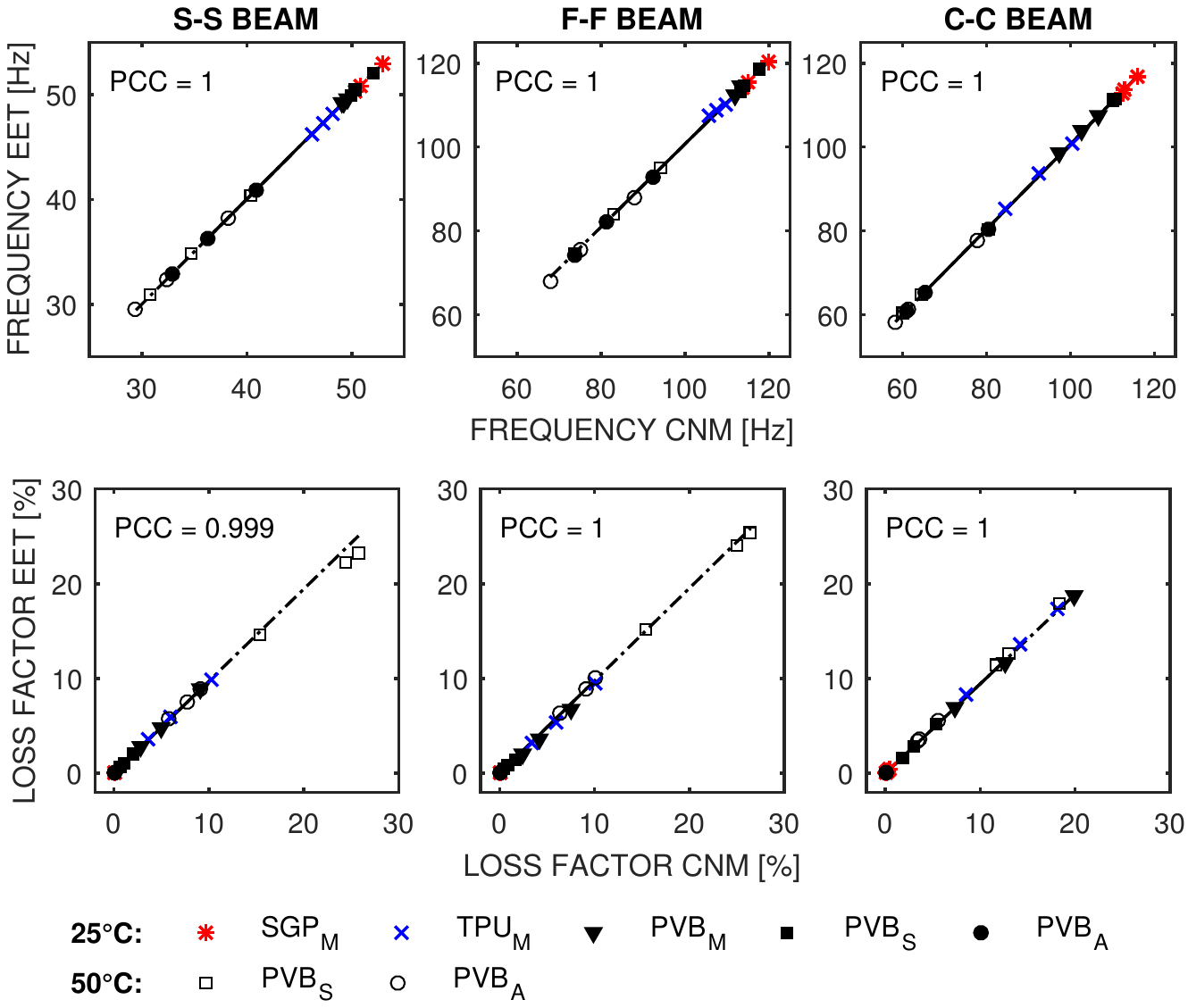}
}
\caption{Quantile-quantile plot for the first mode shape for natural frequencies and loss factors. The response by the adjusted enhanced effective thickness method (EET) is plotted against the reference solution (CNM) along with the Pearson correlation coefficient (PCC).  The Maxwell chain parameters for PVB, TPU, and SGP are taken from \citep{Mohagheghian:2017:QSB} (M), \citep{Shitanoki:2014:PNM} (S), and~\citep{Andreozzi:2014:DTT} (A).}
\label{fig:QQ_SET}
\end{figure}

In analogy to the previous sections, the errors associated with the EET method against the reference solution (CNM) are plotted in~\Fref{fig:e3SET} and the quantile-quantile plots appear in~\Fref{fig:QQ_SET}.


For the simply supported beam, the errors remain below 1$\%$ for frequencies and below the 10$\%$ for loss factors, as in the case of the DET method. For the other boundary conditions, the errors in both, frequencies and loss factors, are lower than in the case of the DET approach. More specifically, they are below 5$\%$ for frequencies and below 22$\%$ for loss factors. The highest errors appear for the free-free beams, the lowest again for the simply-supported beam. This method provides good approximations of natural frequencies for all three boundary conditions and the best estimates of loss factors from all tested simplified methods.  

\subsubsection{Comparison of simplified approaches}

\begin{figure}[h]
\centerline{
\includegraphics[width=110mm]{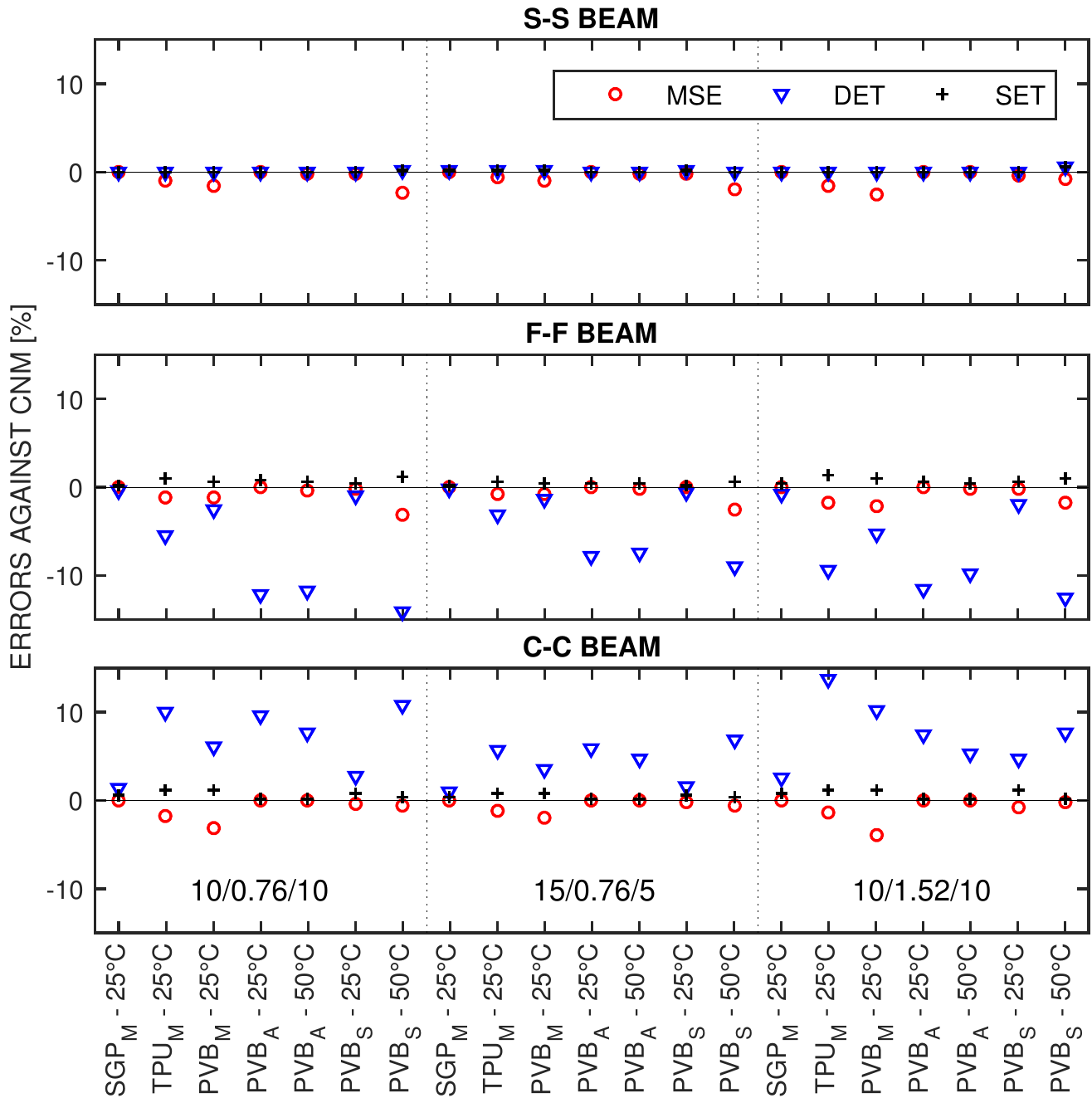}
}
\caption{Summary of errors in natural frequencies obtained by all simplified methods against the reference solution (CNM) corresponding to the first mode shape for all tested cases. Simplified approaches: modal strain energy~(MSE), dynamic effective thickness~(DET), and enhanced effective thickness~(EET) methods.} 
\label{fig:eF1MS}
\end{figure}
\begin{figure}[h]
\centerline{
\includegraphics[width=110mm]{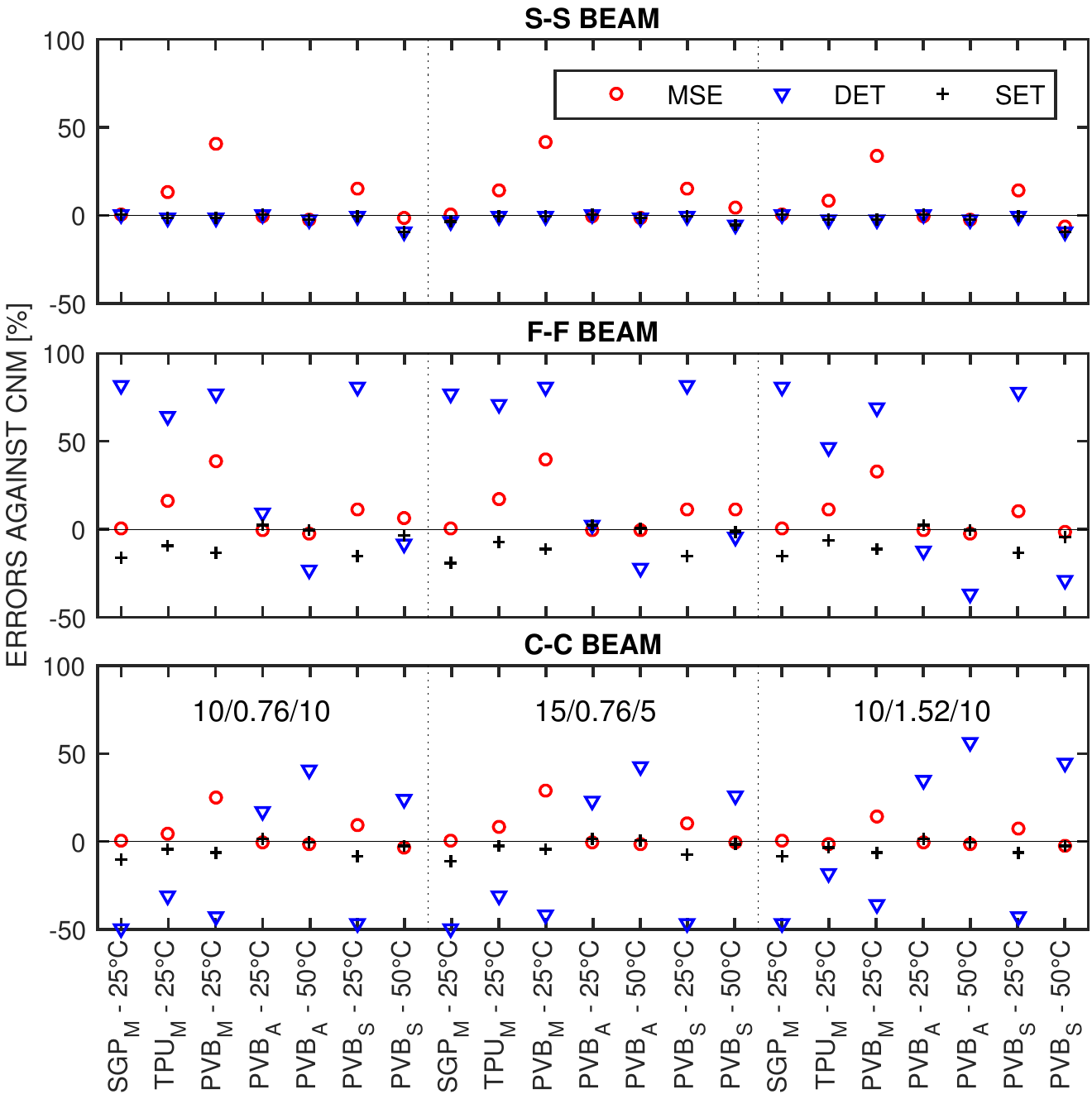}
}
\caption{Summary of errors in loss factors obtained by all simplified methods against the reference solution (CNM) corresponding to the first mode shape for all tested cases. Simplified approaches: modal strain energy~(MSE), dynamic effective thickness~(DET), and enhanced effective thickness~(EET) methods.} 
\label{fig:eL1MS}
\end{figure}

Finally, the errors of both quantities predicted by all three methods are summarized in Figures~\ref{fig:eF1MS} and \ref{fig:eL1MS} for all tested examples for the first mode shape. The present study thus shows that:
\begin{itemize}
\item 
For natural frequencies, modal strain energy and effective thickness methods give very good approximations (below 4$\%$ for MSE and EET, below 15$\%$ for DET). 

\item
The effective thickness approaches provide excellent results for simply-supported laminated glass beams; the error is less than 1\%.

\item
However, all methods predict in some cases the loss factors of laminated glass beams with errors in tens of percent (up to 22$\%$ for EET, 42$\%$ for MSE, and 85$\%$ for DET). Therefore, they can provide only an informative estimate of damping. 

\item
For simply-supported beams, the effective thickness methods deliver the loss factors with errors less than 5\% for the room temperature and below 10\% for the evaluated temperature typical of, e.g., an exterior facade panel in summer.

\item
The algorithm based on the EET provides the best estimates of loss factors from all tested simplified methods.

\item The symmetry or the asymmetry of the cross-section geometry does not influence the level of errors.

\end{itemize}

It is worth mentioning that for the simply-supported beams the enhanced effective method adjusted for dynamics (EET) gives exactly the same results as the original dynamic effective thickness method (DET). We can see from Tables~\ref{tab:psi} and \ref{tab:wn} that shape coefficients are squared wavenumbers for a simply-supported beam $\psi = \wn^2$. If we replaced the used shape coefficient derived in~\Tref{tab:psi} with squared wavenumbers from~\Tref{tab:wn}, we would obtain the same results from both methods for all boundary conditions. A more detailed study of this unexpected connection goes beyond the current work and will be performed separately.

\section{Conclusions}
\label{S:Conc}

Four methods for modal analysis of laminated glass structures were introduced in this paper, i.e., the numerical complex-valued eigensolver based on the Newton method, the real-valued eigensolver complemented with the modal strain energy method, and the original and enhanced dynamic effective thickness method. 
The aim of this paper was to assess the usability of the last three practical methods, comparing their predictions with the complex-valued eigensolver. For the enhanced effective thickness method by~\citeauthor{Galuppi:2012:ETL}, we proposed and presented our extension of the current method for modal analysis of laminated glass. The following conclusions can be made from the present study:
\begin{enumerate}
\item 
Our study underlines the importance of being careful when predicting the damping in laminated glass. The loss factor is a sensitive quantity which affects the errors of approximations provided by simplified methods. In particular, the natural frequencies can be predicted with errors less than 5$\%$ using a suitable method, but the same does not hold for the loss factors.

\item
The level of errors in approximated quantities depends on the boundary conditions for the effective thickness approaches and on the parameters of the generalized Maxwell model for all methods. Material parameters for the interlayers from the literature leads to different natural frequencies and loss factors even for the same type of polymer.

\item The enhanced effective thickness approach adjusted for modal analysis provides approximations of both quantities, natural frequencies and loss factors, with lower errors compared to the other two simplified procedures. 

\end{enumerate}

Simplified methods and approaches reduce computational time and cost and have an important place in the design of engineering structures. Under the above limitations, our study confirms their suitability for modal analysis of laminated glass structures. 

\paragraph{Acknowledgments} 
%
This publication was supported by the Czech Science
Foundation under project No. 16-14770S.

\bibliographystyle{apalike} 
\bibliography{liter.bib}

\appendix
\section{Material properties of glass and interlayers}\label{S:A1}
 In this appendix, we summarize the used material properties of glass and three different polymers (taken from the literature) in~Tables~\ref{T:Glass}, \ref{T:Interlayers}, \ref{T:STP}, and \ref{T:PVB}.
 
\begin{table}[h]
\centering
\begin{tabular}{l c r l}
\hline
\multicolumn{3}{c}{Soda-lime glass}\\
\hline
Density & $\den_1 = \den_3$ & 2,500 & kg.m$^{-3}$\\
Young's modulus of elasticity & $\E_1 = \E_3$ & 72 & GPa\\
Poisson's ratio & $\PR_1 = \PR_3$ & 0.22 & --\\
\hline
\end{tabular}
\caption{Glass properties~\citep{Aenlle:2013:FRLGE}.}
\label{T:Glass}
\end{table}
 
\begin{table}[h]
\centering
\begin{tabular}{l c c c c c}
\hline
& & {SGP$_\textrm{M}$} &
{TPU$_\textrm{M}$} &
{PVB$_\textrm{M}$, PVB$_\textrm{S}$, PVB$_\textrm{A}$} & \\
\hline
Density & {$\den_2$} & 950 & 1070 & 1100 & kg.m$^{-3}$\\
Poisson's ratio & {$\PR_2$} & 0.49 & 0.49 & 0.49 & --\\
\hline
\end{tabular}
\caption{Interlayer properties~\citep{Mohagheghian:2017:QSB}.}
\label{T:Interlayers}
\end{table}

\newpage

\begin{table}[h]
\centering
\begin{tabular}{c c c c c}
\hline
& {SGP$_\textrm{M}$} & {TPU$_\textrm{M}$} & {PVB$_\textrm{M}$} \\
\hline
\multicolumn{5}{c}{Long-term and instantaneous shear moduli}\\
\multicolumn{1}{c}{$\Ginf$} & 1.8 & 1.56 & 0.22 & MPa\\
\multicolumn{1}{c}{$\GO$} & 274.1 & 94.6 & 213.6 & MPa \\
\hline
\multicolumn{5}{c}{Coefficients and relaxation times}\\
$\prony$ &
 \multicolumn{3}{c}{$\Gp/\GO$} & $\tp$ [s] 
 \\
1 & 0.07767 & 0.42077 & 0.39262 & $10^{-6}$  \\
2 & 0.03764 & 0.18113 & 0.19225 & $10^{-5}$ \\
3 & 0.05631 & 0.19280 & 0.20957 & $10^{-4}$ \\
4 & 0.06501 & 0.09969 & 0.12621 & $10^{-3}$ \\
5 & 0.07409 & 0.04750 & 0.05694 & $10^{-2}$ \\
6 & 0.09317 & 0.01928 & 0.01536 & $10^{-1}$ \\
7 & 0.11867 & 0.00903 & 0.00325 & $10^{0}$ \\
8 & 0.20551 & 0.00414 & 0.00103 & $10^{1}$ \\
9 & 0.18131 & 0.00307 & 0.00077 & $10^{2}$ \\
10 & 0.05361 & 0.00230 & 0.00010 & $10^{3}$ \\
11 & 0.01856 & 0.00371 & 0.00029 & $10^{4}$ \\
12 & 0.01180 & 0.00004 & 0.00053 & $10^{5}$ \\
\hline
\end{tabular}
\caption{Parameters of the generalized Maxwell model for SGP$_\textrm{M}$, TPU$_\textrm{M}$, and PVB$_\textrm{M}$~\citep{Mohagheghian:2017:QSB}.}
\label{T:STP}
\end{table}

\newpage

\begin{table}[h]
\centering
\begin{tabular}{c c c c c}
\hline
& \multicolumn{2}{c}{{PVB$_\textrm{S}$}}
& \multicolumn{2}{c}{{PVB$_\textrm{A}$}}
\\
\hline
\multicolumn{5}{c}{Long-term shear moduli}\\
\multicolumn{1}{c}{$\Ginf$} & \multicolumn{2}{c}{0} & \multicolumn{2}{c}{0} \\
\hline
\multicolumn{5}{c}{Shear moduli and relaxation times}\\
$\prony$ &
{$\Gp$} [MPa] & $\tp$ [s] 
&
{$\Gp$} [MPa] & $\tp$ [s] 
 \\
1 &51.25 &4.273$\times10^{-7}$& 0.514628 & 9.51$\times10^{-2}$ \\
2 &31.75 &3.546$\times10^{-6}$& 0.280116 & 4.71$\times10^{-1}$\\
3 &12.80 &1.330$\times10^{-5}$& 0.144282 & 2.72$\times10^{0}$ \\
4 &32.90 &4.279$\times10^{-5}$& 0.086904 & 2.11$\times10^{1}$\\
5 &39.90 &2.984$\times10^{-4}$& 0.076190 & 2.21$\times10^{2}$\\
6 &37.80 &2.170$\times10^{-3}$& 0.092202 & 2.12$\times10^{3}$\\
7 &21.94 &8.274$\times10^{-3}$& 0.098780 & 1.74$\times10^{4}$\\
8 &25.01 &2.937$\times10^{-2}$& 0.085555 & 1.31$\times10^{5}$ \\
9 &27.58 &1.658$\times10^{-1}$& 0.070251 & 1.05$\times10^{6}$\\
10 &11.98&7.774$\times10^{-1}$& 0.107653 & 2.99$\times10^{7}$\\
11 &6.345&3.293$\times10^{0}$\\
12 &2.692&1.698$\times10^{1}$\\
13 &8.718&2.041$\times10^{2}$\\
14 & 0.6969 &3.588$\times10^{4}$\\
\hline
\multicolumn{5}{c}{Parameters for temperature shifting}\\
\multicolumn{1}{c}{$T_0$} & \multicolumn{2}{c}{20.46$\,^\circ\textrm{C}$} & \multicolumn{2}{c}{30$\,^\circ\textrm{C}$} \\
\multicolumn{1}{c}{$C_1$} & \multicolumn{2}{c}{37.30} & \multicolumn{2}{c}{12.5} \\
\multicolumn{1}{c}{$C_2$} & \multicolumn{2}{c}{203.61$\,^\circ\textrm{C}$} & \multicolumn{2}{c}{89$\,^\circ\textrm{C}$} \\
\hline
\end{tabular}
\caption{Parameters of the generalized Maxwell model for PVB$_\textrm{S}$~\citep{Shitanoki:2014:PNM} and PVB$_\textrm{A}$~\citep{Andreozzi:2014:DTT}.}
\label{T:PVB}
\end{table}

\newpage
\newpage
\section{Numerical aspects}
\label{S:A_NA}

The finite element discretization is briefly presented in this section, following the exposition by~\citet{rikards1993finite}. Our study deals with a laminated glass beam made of three layers. After discretization, we have 18 unknowns per element -- 9 unknowns per cross section, recall~\Fref{fig:superelement}. Therefore, the kinematics of the three-layer element is specified by the vector of nodal displacements and rotations
\begin{equation*}
\Mue^{\textnormal{full}} = 
\begin{array}[t]{cccccccccc}
\bigl[
\uL{1} & \wL{1} & \rotL{1} & \uP{1} & \wP{1} & \rotP{1} &   \uL{2} & \wL{2} & \rotL{2} & 
\vspace{1mm}\\
 & \uP{2} & \wP{2} & \rotP{2} & \uL{3} & \wL{3} & \rotL{3} & \uP{3} & \wP{3} & \rotP{3} 
\bigr]\trn.
\end{array}
\end{equation*}

Using the Timoshenko beam theory, we express 8 inter-layer compatibility conditions corresponding to perfect horizontal and vertical adhesion at each inter-layer interface as
\begin{subequations}
\begin{align}
\uL{1} + \frac{h_1}{2} \rotL{1} &= \uL{2} - \frac{h_2}{2} \rotL{2} ,& 
\uP{1} + \frac{h_1}{2} \rotP{1} &= \uP{2} - \frac{h_2}{2} \rotP{2},\nonumber \\ 
\uL{2} + \frac{h_2}{2} \rotL{2} &= \uL{3} - \frac{h_3}{2} \rotL{3},& 
\uP{2} + \frac{h_2}{2} \rotP{2} &= \uP{3} - \frac{h_3}{2} \rotP{3}, \nonumber \\
\wL{1} &= \wL{2},& 
\wP{1} &= \wP{2},\nonumber \\ 
\wL{2} &= \wL{3},& 
\wP{2} &= \wP{3}. \nonumber
\end{align}
\end{subequations}
Therefore, we decompose the vector of the generalized nodal displacements as follows
\begin{subequations}
\begin{align}
\M{u}_{e}
&=
\begin{array}[t]{ccccccccccc}
\bigl[
\uL{1} & \wL{1} & \rotL{1} & \uL{3} &  \rotL{3} & 
\uP{1} & \wP{1} & \rotP{1} & \uP{3} & \rotP{3}
\bigr]\trn,
\end{array}
\nonumber
\\
\M{u}_{e}^{\textnormal{slave}}
&=
\begin{array}[t]{ccccccccc}
\bigl[
\uL{2} & \wL{2} & \rotL{2} & \wL{3} &
 \uP{2} & \wP{2} & \rotP{2} &  \wP{3}
\bigr]\trn,
\end{array}
\nonumber
\end{align}
\end{subequations}
where $\M{u}_{e}$ are the independent unknowns and $\M{u}_{e}^{\textnormal{slave}}$ are the dependent ones.
Then, the compatibility conditions can be written in the compact form
\begin{equation*}
\M{u}_{e}^{\textnormal{full}} = \M{T}_e \M{u}_{e}
\end{equation*}
with the transformation matrix
\begin{equation*}
\M{T}_e =
\left[
\begin{array}{c;{1pt/3pt}c;{1pt/3pt}c;{1pt/3pt}c;{1pt/3pt}c;{1pt/3pt}c;{1pt/3pt}c;{1pt/3pt}c;{1pt/3pt}c;{1pt/3pt}c}
1 & & & & & & & &\\ \hdashline[1pt/3pt]
& 1 & & & & & & &\\ \hdashline[1pt/3pt]
& & 1 & & & & & &\\ \hdashline[1pt/3pt]
& & & & & 1 & & & \\ \hdashline[1pt/3pt]
& & & & & & 1 & & \\ \hdashline[1pt/3pt]
& & & & & & & 1 & \\ \hdashline[1pt/3pt]
\frac{1}{2} & & \frac{h_1}{4} & \frac{1}{2} & -\frac{h_3}{4} & & & & &\\ \hdashline[1pt/3pt]
& 1 & & & & & & &\\ \hdashline[1pt/3pt]
-\frac{1}{h_2} & & -\frac{h_1}{2h_2} & \frac{1}{2h_2} & -\frac{h_3}{2 h_2} & & & & & \\ \hdashline[1pt/3pt]
& & & & & \frac{1}{2} & & \frac{h_1}{4} & \frac{1}{2} & -\frac{h_3}{4}\\ \hdashline[1pt/3pt]
& & & & & & 1 & & &\\ \hdashline[1pt/3pt]
& & & & &-\frac{1}{h_2} & & -\frac{h_1}{2h_2} & \frac{1}{2h_2} & -\frac{h_3}{2 h_2}\\ \hdashline[1pt/3pt]
& & & 1 & & & & & &\\ \hdashline[1pt/3pt]
& 1 & & & & & & & &\\ \hdashline[1pt/3pt]
& & & & 1 & & & & &\\ \hdashline[1pt/3pt]
& & & & & & & & 1 &\\ \hdashline[1pt/3pt]
& & & & & & 1 & & &\\ \hdashline[1pt/3pt]
& & & & & & & & & 1 
\end{array}
\right].
\end{equation*}

Using this elimination of unknowns, the element mass matrix $\MM_e$, the initial stiffness matrix $\M{K}_{0,e}$ and the matrix $\M{K}_{\textnormal{c},e}$, recall ~\Eref{eq:SCGEQ}, corresponding to the independent generalized nodal displacements follow from
\begin{align*}
\MM_e = \M{T}\trn \MM_e^{\textnormal{full}} \M{T}, && 
\M{K}_{0,e} = \M{T}\trn \M{K}_{0,e}^{\textnormal{full}} \M{T}, && 
\M{K}_{\textnormal{c},e} = \M{T}\trn \M{K}_{\textnormal{c},e}^{\textnormal{full}} \M{T}
.
\end{align*}
Therefore, they are $10 \times 10$ matrices
derived from the original $18 \times 18$ one. We use the consistent mass matrix and the stiffness matrices are derived using the selective integration scheme to avoid shear locking.

\end{document}

%% file: format.tex
\newcommand{\Fref}[1]{Figure~\ref{#1}}
\newcommand{\Eref}[1]{Eq.~\eqref{#1}}

\newcommand{\Sref}[1]{Section~\ref{#1}}
\newcommand{\Tref}[1]{Table~\ref{#1}}

\newcommand{\del}[2]{\mbox{$\displaystyle\frac{#1}{#2}$}}

\newcommand{\ppd}[2]{\del{\partial {#1}}{\partial{#2}}}

\newcommand{\M}[1]{{\boldsymbol #1}}
\newcommand{\trn}{{\sf ^T}}  
\newcommand{\ite}[1]{{^{#1}}}
\newcommand{\step}{k}
\newcommand{\freq}{f_\text{Hz}} 
\newcommand{\loss}{\eta} 

\newcommand{\freqa}{\omega} 
\newcommand{\emode}{U} 
\newcommand{\Memode}{\M{\emode}}
\newcommand{\den}{\rho} 
\newcommand{\E}{E} 
\newcommand{\PR}{\nu} 
\newcommand{\G}{G} 
\newcommand{\Gc}{\G_{2}} 
\newcommand{\prony}{p} 
\newcommand{\Gp}{\G_{2,\prony}} 
\newcommand{\tp}{\theta_{\prony}} 
\newcommand{\Ginf}{\G_{2,\infty}} 
\newcommand{\GO}{\G_{2,0}}
\newcommand{\Gfr}{\G_{2,\freqa}} 
\newcommand{\MK}{\M{K}} 
\newcommand{\MKO}{\M{K}_{0}} 
\newcommand{\MKfr}{\M{K}_{\freqa}} 
\newcommand{\MKconst}{\M{K}_{\textnormal{c}}} 
\newcommand{\MM}{\M{M}} 
\newcommand{\MT}{\M{T}} 
\newcommand{\rAp}{\textrm{ap}} 
\newcommand{\rO}{\textrm{0}}
\newcommand{\rE}{\textrm{ef}}
\newcommand{\wn}{\beta}
\newcommand{\Mue}{\M{u}_e}
\newcommand{\uL}[1]{u_{#1,\textnormal{L}}}
\newcommand{\uP}[1]{u_{#1,\textnormal{R}}}
\newcommand{\wL}[1]{w_{#1,\textnormal{L}}}
\newcommand{\wP}[1]{w_{#1,\textnormal{R}}}
\newcommand{\rotL}[1]{\varphi_{#1,\textnormal{L}}}
\newcommand{\rotP}[1]{\varphi_{#1,\textnormal{R}}}

%% file: sandwich.pdf_tex
\begingroup%
  \makeatletter%
  \providecommand\color[2][]{%
    \errmessage{(Inkscape) Color is used for the text in Inkscape, but the package 'color.sty' is not loaded}%
    \renewcommand\color[2][]{}%
  }%
  \providecommand\transparent[1]{%
    \errmessage{(Inkscape) Transparency is used (non-zero) for the text in Inkscape, but the package 'transparent.sty' is not loaded}%
    \renewcommand\transparent[1]{}%
  }%
  \providecommand\rotatebox[2]{#2}%
  \ifx\svgwidth\undefined%
    \setlength{\unitlength}{396.8503937bp}%
    \ifx\svgscale\undefined%
      \relax%
    \else%
      \setlength{\unitlength}{\unitlength * \real{\svgscale}}%
    \fi%
  \else%
    \setlength{\unitlength}{\svgwidth}%
  \fi%
  \global\let\svgwidth\undefined%
  \global\let\svgscale\undefined%
  \makeatother%
  \begin{picture}(1,0.21428571)%
    \put(0,0){\includegraphics[width=\unitlength,page=1]{sandwich.pdf}}%
    \put(0.8612387,0.14176897){\color[rgb]{0,0,0}\makebox(0,0)[lb]{\smash{glass}}}%
    \put(0.79607277,0.02107566){\color[rgb]{0,0,0}\makebox(0,0)[rb]{\smash{polymer}}}%
    \put(0,0){\includegraphics[width=\unitlength,page=2]{sandwich.pdf}}%
    \put(0.86137886,0.07790353){\color[rgb]{0,0,0}\makebox(0,0)[lb]{\smash{glass}}}%
    \put(0,0){\includegraphics[width=\unitlength,page=3]{sandwich.pdf}}%
    \put(0.89843848,0.19254586){\color[rgb]{0,0,0}\makebox(0,0)[lb]{\smash{\textit{$y$}}}}%
    \put(0.7942949,0.09012604){\color[rgb]{0,0,0}\makebox(0,0)[lb]{\smash{\textit{$z$}}}}%
    \put(0.02331933,0.08963579){\color[rgb]{0,0,0}\makebox(0,0)[lb]{\smash{\textit{$z$}}}}%
    \put(0.12514181,0.19215676){\color[rgb]{0,0,0}\makebox(0,0)[lb]{\smash{\textit{$x$}}}}%
    \put(0.89836856,0.02575051){\color[rgb]{0,0,0}\makebox(0,0)[b]{\smash{\textit{$b$}}}}%
    \put(0.33408655,0.0253224){\color[rgb]{0,0,0}\makebox(0,0)[b]{\smash{\textit{$l$}}}}%
    \put(0,0){\includegraphics[width=\unitlength,page=4]{sandwich.pdf}}%
    \put(0.67437063,0.13995318){\color[rgb]{0,0,0}\makebox(0,0)[lb]{\smash{}}}%
    \put(0.71029178,0.1396007){\color[rgb]{0,0,0}\makebox(0,0)[lb]{\smash{\textit{$h_1$}}}}%
    \put(0.71029178,0.07322562){\color[rgb]{0,0,0}\makebox(0,0)[lb]{\smash{\textit{$h_3$}}}}%
    \put(0.71029178,0.11037816){\color[rgb]{0,0,0}\makebox(0,0)[lb]{\smash{\textit{$h_2$}}}}%
    \put(0,0){\includegraphics[width=\unitlength,page=5]{sandwich.pdf}}%
  \end{picture}%
\endgroup%

%% file: GMM.pdf_tex
\begingroup%
  \makeatletter%
  \providecommand\color[2][]{%
    \errmessage{(Inkscape) Color is used for the text in Inkscape, but the package 'color.sty' is not loaded}%
    \renewcommand\color[2][]{}%
  }%
  \providecommand\transparent[1]{%
    \errmessage{(Inkscape) Transparency is used (non-zero) for the text in Inkscape, but the package 'transparent.sty' is not loaded}%
    \renewcommand\transparent[1]{}%
  }%
  \providecommand\rotatebox[2]{#2}%
  \ifx\svgwidth\undefined%
    \setlength{\unitlength}{283.46456693bp}%
    \ifx\svgscale\undefined%
      \relax%
    \else%
      \setlength{\unitlength}{\unitlength * \real{\svgscale}}%
    \fi%
  \else%
    \setlength{\unitlength}{\svgwidth}%
  \fi%
  \global\let\svgwidth\undefined%
  \global\let\svgscale\undefined%
  \makeatother%
  \begin{picture}(1,0.25)%
    \put(0.09,0.06){\color[rgb]{0,0,0}\makebox(0,0)[lb]{\smash{}}}%
    \put(0.10082643,0.14904714){\color[rgb]{0,0,0}\makebox(0,0)[rb]{\smash{$G_{2,\infty}$}}}%
    \put(0,0){\includegraphics[width=\unitlength,page=1]{GMM.pdf}}%
    \put(0.24508051,0.07332977){\color[rgb]{0,0,0}\makebox(0,0)[rb]{\smash{$\mu_{1}$}}}%
    \put(0.24500576,0.15172891){\color[rgb]{0,0,0}\makebox(0,0)[rb]{\smash{$G_{2,1}$}}}%
    \put(0.38895033,0.15172891){\color[rgb]{0,0,0}\makebox(0,0)[rb]{\smash{$G_{2,2}$}}}%
    \put(0.38902506,0.07332977){\color[rgb]{0,0,0}\makebox(0,0)[rb]{\smash{$\mu_{2}$}}}%
    \put(0.53296959,0.07332977){\color[rgb]{0,0,0}\makebox(0,0)[rb]{\smash{$\mu_{3}$}}}%
    \put(0.53289484,0.15172891){\color[rgb]{0,0,0}\makebox(0,0)[rb]{\smash{$G_{2,3}$}}}%
    \put(0.85583315,0.07332977){\color[rgb]{0,0,0}\makebox(0,0)[rb]{\smash{$\mu_{P}$}}}%
    \put(0.85575847,0.15172891){\color[rgb]{0,0,0}\makebox(0,0)[rb]{\smash{$G_{2,P}$}}}%
    \put(0,0){\includegraphics[width=\unitlength,page=2]{GMM.pdf}}%
  \end{picture}%
\endgroup%

%% file: superelement.pdf_tex
\begingroup%
  \makeatletter%
  \providecommand\color[2][]{%
    \errmessage{(Inkscape) Color is used for the text in Inkscape, but the package 'color.sty' is not loaded}%
    \renewcommand\color[2][]{}%
  }%
  \providecommand\transparent[1]{%
    \errmessage{(Inkscape) Transparency is used (non-zero) for the text in Inkscape, but the package 'transparent.sty' is not loaded}%
    \renewcommand\transparent[1]{}%
  }%
  \providecommand\rotatebox[2]{#2}%
  \ifx\svgwidth\undefined%
    \setlength{\unitlength}{396.8503937bp}%
    \ifx\svgscale\undefined%
      \relax%
    \else%
      \setlength{\unitlength}{\unitlength * \real{\svgscale}}%
    \fi%
  \else%
    \setlength{\unitlength}{\svgwidth}%
  \fi%
  \global\let\svgwidth\undefined%
  \global\let\svgscale\undefined%
  \makeatother%
  \begin{picture}(1,0.35714286)%
    \put(0,0){\includegraphics[width=\unitlength,page=1]{superelement.pdf}}%
    \put(0.87660386,0.22737579){\color[rgb]{0,0,0}\makebox(0,0)[lb]{\smash{glass}}}%
    \put(0,0){\includegraphics[width=\unitlength,page=2]{superelement.pdf}}%
    \put(0.87660386,0.07022487){\color[rgb]{0,0,0}\makebox(0,0)[lb]{\smash{glass}}}%
    \put(0,0){\includegraphics[width=\unitlength,page=3]{superelement.pdf}}%
    \put(0.15567052,0.22535008){\color[rgb]{0,0,0}\makebox(0,0)[lb]{\smash{\textit{$z$}}}}%
    \put(0.26127276,0.32787105){\color[rgb]{0,0,0}\makebox(0,0)[lb]{\smash{\textit{$x$}}}}%
    \put(0.50050937,0.33205534){\color[rgb]{0,0,0}\makebox(0,0)[b]{\smash{\textit{$l_e$}}}}%
    \put(0,0){\includegraphics[width=\unitlength,page=4]{superelement.pdf}}%
    \put(0.67437063,0.13995318){\color[rgb]{0,0,0}\makebox(0,0)[lb]{\smash{}}}%
    \put(0.78659881,0.2277965){\color[rgb]{0,0,0}\makebox(0,0)[lb]{\smash{\textit{$h_1$}}}}%
    \put(0.78659881,0.07064558){\color[rgb]{0,0,0}\makebox(0,0)[lb]{\smash{\textit{$h_3$}}}}%
    \put(0.78659881,0.14922158){\color[rgb]{0,0,0}\makebox(0,0)[lb]{\smash{\textit{$h_2$}}}}%
    \put(0,0){\includegraphics[width=\unitlength,page=5]{superelement.pdf}}%
    \put(0.42472097,0.20711343){\color[rgb]{1,0,0}\makebox(0,0)[lb]{\smash{\textit{$\uL{1}$}}}}%
    \put(0.27949481,0.19786473){\color[rgb]{1,0,0}\makebox(0,0)[lb]{\smash{\textit{$\rotL{1}$}}}}%
    \put(0.37053337,0.17849775){\color[rgb]{1,0,0}\makebox(0,0)[lb]{\smash{\textit{$\wL{1}$}}}}%
    \put(0,0){\includegraphics[width=\unitlength,page=6]{superelement.pdf}}%
    \put(0.42472097,0.12854222){\color[rgb]{0.30196078,0.30196078,0.30196078}\makebox(0,0)[lb]{\smash{\textit{$\uL{2}$}}}}%
    \put(0.27949481,0.11929352){\color[rgb]{0.30196078,0.30196078,0.30196078}\makebox(0,0)[lb]{\smash{\textit{$\rotL{2}$}}}}%
    \put(0.37053337,0.09992654){\color[rgb]{0.30196078,0.30196078,0.30196078}\makebox(0,0)[lb]{\smash{\textit{$\wL{2}$}}}}%
    \put(0,0){\includegraphics[width=\unitlength,page=7]{superelement.pdf}}%
    \put(0.42472097,0.04997079){\color[rgb]{1,0,0}\makebox(0,0)[lb]{\smash{\textit{$\uL{3}$}}}}%
    \put(0.27949481,0.04072209){\color[rgb]{1,0,0}\makebox(0,0)[lb]{\smash{\textit{$\rotL{3}$}}}}%
    \put(0.37053337,0.02135511){\color[rgb]{0.30196078,0.30196078,0.30196078}\makebox(0,0)[lb]{\smash{\textit{$\wL{3}$}}}}%
    \put(0,0){\includegraphics[width=\unitlength,page=8]{superelement.pdf}}%
    \put(0.87552572,0.14880088){\color[rgb]{0,0,0}\makebox(0,0)[lb]{\smash{polymer\\ }}}%
    \put(0,0){\includegraphics[width=\unitlength,page=9]{superelement.pdf}}%
    \put(0.71043522,0.20711343){\color[rgb]{1,0,0}\makebox(0,0)[lb]{\smash{\textit{$\uP{1}$}}}}%
    \put(0.56520912,0.19786473){\color[rgb]{1,0,0}\makebox(0,0)[lb]{\smash{\textit{$\rotP{1}$}}}}%
    \put(0.65624766,0.17849775){\color[rgb]{1,0,0}\makebox(0,0)[lb]{\smash{\textit{$\wP{1}$}}}}%
    \put(0,0){\includegraphics[width=\unitlength,page=10]{superelement.pdf}}%
    \put(0.71043522,0.12854222){\color[rgb]{0.30196078,0.30196078,0.30196078}\makebox(0,0)[lb]{\smash{\textit{$\uP{2}$}}}}%
    \put(0.56520912,0.11929352){\color[rgb]{0.30196078,0.30196078,0.30196078}\makebox(0,0)[lb]{\smash{\textit{$\rotP{2}$}}}}%
    \put(0.65624766,0.09992654){\color[rgb]{0.30196078,0.30196078,0.30196078}\makebox(0,0)[lb]{\smash{\textit{$\wP{2}$}}}}%
    \put(0,0){\includegraphics[width=\unitlength,page=11]{superelement.pdf}}%
    \put(0.71043522,0.04997079){\color[rgb]{1,0,0}\makebox(0,0)[lb]{\smash{\textit{$\uP{3}$}}}}%
    \put(0.56520912,0.04072209){\color[rgb]{1,0,0}\makebox(0,0)[lb]{\smash{\textit{$\rotP{3}$}}}}%
    \put(0.65624766,0.02135511){\color[rgb]{0.30196078,0.30196078,0.30196078}\makebox(0,0)[lb]{\smash{\textit{$\wP{3}$}}}}%
    \put(0.2313609,0.07970625){\color[rgb]{1,0,0}\makebox(0,0)[rb]{\smash{master DOFs}}}%
    \put(0.2313609,0.03599025){\color[rgb]{0.30196078,0.30196078,0.30196078}\makebox(0,0)[rb]{\smash{slave DOFs}}}%
  \end{picture}%
\endgroup%